\documentclass[11pt,letterpaper]{article}

\usepackage[title]{appendix}
\usepackage[english]{babel}
\usepackage[utf8]{inputenc}
\usepackage{amsmath}
\usepackage{amsthm}
\usepackage{mathtools}
\usepackage{amssymb}
\usepackage{upgreek}
\usepackage{graphicx}
\usepackage{geometry}
\usepackage{cite}
\usepackage{authblk}
\usepackage{color}
\usepackage{soul}
\usepackage{xcolor}
\usepackage{caption}
\usepackage{subcaption}
\usepackage{makecell}

\usepackage[linesnumbered,ruled,vlined]{algorithm2e}
\SetKwInput{KwInput}{Input}                
\SetKwInput{KwOutput}{Output}              

\usepackage{float}
\usepackage{booktabs}
\usepackage{tabu} 
\usepackage{xcolor}
\usepackage{dcolumn}
\newcolumntype{A}{D{.}{.}{2.3}}
\usepackage[hidelinks]{hyperref}
\newtheorem{prop}{Proposition}

\newcommand{\EQ}{\begin{equation}\begin{aligned}}
\newcommand{\EN}{\end{aligned}\end{equation}}

\newcommand{\GQ}{\begin{equation}\begin{gathered}}
\newcommand{\GN}{\end{gathered}\end{equation}}

\DeclareMathOperator*{\argmax}{arg\,max}

\title{Optimal Design of Frame Structures with Mixed Categorical and Continuous Design Variables Using the Gumbel-Softmax Method\thanks{This is a preprint of an article published in \emph{Structural and Multidisciplinary Optimization}. The final authenticated version is available online at: \href{https://doi.org/10.1007/s00158-024-03745-7}{https://doi.org/10.1007/s00158-024-03745-7}.}}

\author[a]{\small Mehran Ebrahimi\footnote{\textit{E-mail address: }mehran.ebrahimi@autodesk.com}\textsuperscript{,}}
\author[a]{\small Hyunmin Cheong}
\author[a]{\small Pradeep Kumar Jayaraman}
\author[a]{\small Farhad Javid}

\affil[a]{\footnotesize Autodesk Research, 661 University Avenue, Toronto, ON M5G 1M1, Canada}

\date{}

\begin{document}
\maketitle

\begin{abstract}
\noindent
In optimizing real-world structures, due to fabrication or budgetary restraints, the design variables may be restricted to a set of standard engineering choices. Such variables, commonly called categorical variables, are discrete and unordered in essence, precluding the utilization of gradient-based optimizers for the problems containing them. In this paper, incorporating the Gumbel-Softmax (GSM) method, we propose a new gradient-based optimizer for handling such variables in the optimal design of large-scale frame structures. The GSM method provides a means to draw differentiable samples from categorical distributions, thereby enabling sensitivity analysis for the variables generated from such distributions. The sensitivity information can greatly reduce the computational cost of traversing high-dimensional and discrete design spaces in comparison to employing gradient-free optimization methods. In addition, since the developed optimizer is gradient-based, it can naturally handle the simultaneous optimization of categorical and continuous design variables. Through three numerical case studies, different aspects of the proposed optimizer are studied and its advantages over population-based optimizers, specifically a genetic algorithm, are demonstrated.
\end{abstract}


\section{Introduction}
\label{sec:introduction}
The application of frame structures composed of interconnected beams and columns is ubiquitous in various fields of engineering making their simulation and optimization attractive subjects of interest in both academia and industry \cite{mcguire1982matrix, huang1997optimal, kaveh2012charged, ebrahimi2021low}. Commonly, the finite element (FE) analysis of frame structures is performed by modeling their components with beam elements which are capable of undergoing longitudinal (axial), transverse (bending) and torsional deformations, unlike truss elements that can handle only the first one. Accurately capturing these deformation modes and their interactions makes the simulation of frame structures computationally more demanding than that for truss structures \cite{ebrahimi2021low}. The difference becomes particularly more noticeable and exceedingly cumbersome in the optimal design of large-scale frame structures which has led to considerably fewer extensive investigations of this topic for such structures in the literature. 

The focus of the present paper is the optimal design of frame structures with mixed categorical and continuous design variables. Categorical design variables are inherently discrete and their values belong to an unordered set of available choices. Examples of such variables in frame structures include beam cross-sections (profiles) and materials. In real-world applications, only a limited number of cross-sectional profiles and material choices may be available to design a structure due to various limitations such as manufacturing process and cost. For instance, the beam material may only be steel or aluminum, and its cross-section may only be I-profile or T-profile with limited options for their geometrical attributes. Therefore, for practical purposes, it is crucial to distinguish between categorical and continuous design variables, as the latter ones are free to take any value within their bounds and are more straightforward to handle.

The challenges associated with considering categorical design variables are not exclusive to frame structures and have been explored significantly in other application areas \cite{chapman1994genetic, angelov2003automatic, lund2005structural, cheong2019configuration, piacentini2020multi}. Of particular relevance to this paper are truss structures and the techniques developed for handling categorical variables therein. A worthwhile review of the available techniques for truss structures can be found in \cite{stolpe2016truss}. Although frame and truss structures behave differently, they bear a close resemblance in the way they are designed. Thus, the methodology and arguments presented in this paper are directly applicable to truss structures as well.
 
Broadly, existing optimization techniques applied to structural problems with categorical variables can be classified into gradient-free and gradient-based schemes. The majority of methods developed for these problems utilize gradient-free optimizers such as the branch-and-bound \cite{huang1997optimal, stolpe2005design}, genetic algorithm (GA) \cite{jenkins1992plane, kaveh2004size}, simulated annealing \cite{park2002optimization, kripka2004discrete} and particle swarm optimization \cite{kaveh2009particle, li2009heuristic}, among others. It is well-known that due to the combinatorial nature of problems with categorical design variables, the computational cost associated with exploring their design space, which typically grows exponentially with the number of variables, becomes intractable for large-scale problems using the gradient-free optimization routines \cite{yates1982complexity, fister2013brief, shahabsafa2018novel}. Hence, most cases studied in the literature involve structures with a small number of components and a few (predominantly two) categorical choices per element (component). 

Employing gradient-based optimization techniques can be considered as a remedy to alleviate the scalability issue \cite{fister2013brief, stolpe2016truss, ebrahimi2019design}. However, as the categorical variables are discrete and more importantly unordered, computing the gradients of objective and constraint functions with respect to these variables is problematic. There are hence rare precedents of incorporating gradient-based optimizers in the optimal design of beam and truss structures with categorical design variables \cite{yan2016concurrent, krogh2017gradient, duan2019discrete, barjhoux2020bi}. The proposals in this area apply some form of approximation to compute the function gradients with respect to the categorical variables (e.g., relaxation of discrete to continuous variables \cite{krogh2017gradient}, which only makes sense if the variables are ordered) along with rounding techniques to enforce the discreteness of the categorical variables. In another body of work \cite{yan2016concurrent, duan2019discrete}, the weights associated with each categorical choice, in their case material types, are treated as design variables. Subsequently, in \cite{yan2016concurrent} the Heaviside function has been applied to penalize the intermediate weight values to a particular categorical choice.

One can think of several scenarios where these approaches fall short and a more robust technique is desired. For instance, if cross-sectional profiles are design variables in a problem, say I-, T- and U-profiles, it is mathematically challenging (if not impossible) to interpret the meaning of the gradient of a function with respect to these profiles in a direct sense. One possible solution to this problem is to work with the attributes of categorical variables. For example, instead of operating directly on the cross-sectional profiles, their areas can be taken as design variables. However, this becomes troublesome when categorical variables have multiple attributes influencing the optimization progress in a conflicting manner, such as for a cantilever beam under gravity where its cross-sectional area and its second moment of area can affect the beam's deflection in opposing ways. Furthermore, relaxing the discrete variables to continuous ones---inspired by the SIMP method for the topology optimization of solid structures \cite{bendsoe2003topology}---is often limited to problems wherein categorical variables are ordered or have only two choices.

In this paper, we propose a technique for resolving the aforementioned issues in computing the gradients with respect to categorical variables, thus enabling the use of gradient-based optimizers in structural design problems. This novel optimization scheme is founded upon two main ideas. Firstly, we propose \emph{reparametrizing} the categorical design variables and taking the probability of using them in the structure as their corresponding design variables. For instance, if a categorical variable is a beam's cross-section chosen from I-, T- and U-profiles, we propose taking the probabilities of having I-, T- and U-profiles for that beam as its corresponding design variables. As a result, instead of discrete variables, the optimizer deals with continuous ones. Consequently, the optimization functions must be reparameterized in terms of the probabilities and their sensitivity analysis must be carried out with respect to these probabilities, the details of which are presented in the forthcoming sections. Obviously, defining functions in terms of probabilities makes them probabilistic (i.e., stochastic) and in order to compute their values a sampling process is involved, which is in general nondifferentiable. Therefore, secondly, we propose using the Gumbel-Softmax (GSM) technique for making the sampling process differentiable \cite{jang2016categorical, maddison2017concrete}. 

The GSM method is a relaxed version of the original Gumbel-Max (GM) method \cite{gumbel1954statistical} that provides a simple mechanism for drawing differentiable samples from a categorical probability distribution parameterized by the unnormalized log-probabilities of the classes (choices) of that categorical variable \cite{huijben2022review}. In recent years, since the seminal papers \cite{jang2016categorical} and \cite{maddison2017concrete}, the GSM method and its variants have garnered significant attention in the machine learning community (e.g., \cite{cesa2017boltzmann, gu2018neural, yang2019modeling, kool2019stochastic}).  The intrinsic commonalities between combinatorial problems in machine learning and (structural) engineering applications prompted us to explore the utilization of the GSM technique in frame structures. To the best of the authors' knowledge, this study is the first of its kind to employ the GSM method in optimization problems involving mixed continuous and categorical design variables.

The remainder of the present paper is organized as follows. Section \ref{sec:problemdef} introduces the general problem formulation and the notations involved. In Section \ref{sec:gumbel}, we present the GSM method; the pivotal ingredient of the proposed optimization technique in this study. Then, in Section \ref{sec:method}, the details of the optimizer are laid out. Three numerical examples are provided in Section \ref{sec:case} to assess the performance of the developed optimizer and discuss its various aspects. Finally, the paper is concluded by making a few remarks about our proposal in this article and potential considerations for further improvement. 

\section{Problem statement}
\label{sec:problemdef}
The optimal design problem of a frame structure can be posed as follows. Given a frame structure, find the optimum values of continuous design variables as well as optimum choices for the categorical design variables such that the target structural performance function is achieved and the governing linear elasticity equations due to a FE discretization and imposed constraints are satisfied. Accordingly, the topology of the structure does not change throughout the optimization routine. Denoting $\boldsymbol{x} := [x_1, \cdots, x_{n_x} ]^T $ and $\boldsymbol{c} := [c_1, \cdots, c_{n_c}]^T$ as the vectors of continuous and categorical design variables, respectively, the optimization problem mathematically reads

\begin{alignat}{3}\label{eq:optproblem1}
    \min_{\boldsymbol{x}, \boldsymbol{c}} \quad &J \left( \boldsymbol{u}(\boldsymbol{x}, \boldsymbol{c}), \boldsymbol{x}, \boldsymbol{c} \right)  \\
    \text{subject to} \quad &\mathbf{K}(\boldsymbol{x}, \boldsymbol{c}) \: \boldsymbol{u}(\boldsymbol{x}, \boldsymbol{c}) = \boldsymbol{f}(\boldsymbol{x}, \boldsymbol{c}), \nonumber \\
    &\boldsymbol{g}\left( \boldsymbol{u}(\boldsymbol{x}, \boldsymbol{c}), \boldsymbol{x}, \boldsymbol{c} \right) \leq 0, \nonumber \\
    &lb_i \leq x_i \leq ub_i,\quad \quad &&i=1,\cdots, \nonumber n_x, \\
    &c_i \in \{1,\cdots,N_i \}, \quad \quad &&i=1,\cdots, \nonumber n_c.
\end{alignat}

In this equation, $J$ is the scalar-valued objective function, $\boldsymbol{g} \in \mathbb{R}^{n_g}$ represents the vector of optimization constraint functions, $\mathbf{K} \in \mathbb{R}^{n_u \times n_u}$ indicates the structure's stiffness matrix, $\boldsymbol{f} \in \mathbb{R}^{n_u}$ specifies the external load vector, and $\boldsymbol{u} \in \mathbb{R}^{n_u}$ denotes the vector of nodal displacements (state variables). Also, $lb_i$ and $ub_i, \: i=1,\cdots, n_x,$ refer to the lower and upper bounds of the continuous variable $x_i$, respectively. Furthermore, $N_i, \: i=1,\cdots, n_c$, is the number of choices available for the categorical variable $c_i$. Therefore, without loss of generality, we assume the existing choices for each $c_i, \: i=1,\cdots, n_c,$ are indexed as $\{1,\cdots,N_i \}$. 

Depending on the problem at hand and the capabilities of the FE solver, $J$ and $\boldsymbol{g}$ can be linear or nonlinear functions of their arguments. Other than continuity and (at least first order) differentiability, we make no further assumptions on $J$ and $\boldsymbol{g}$. Given a structure, some examples of these functions include mass, compliance, displacement, von Mises stress and natural frequencies. Since the response of a structure is affected by the values of the design variables, the nodal displacements $\boldsymbol{u}$ are implicit functions of $\boldsymbol{x}$ and $\boldsymbol{c}$. As mentioned in the introduction, we aim to use gradient-based optimizers to solve the problem in \eqref{eq:optproblem1}. However, due to the challenges elaborated earlier, direct computation of $\nabla_{\boldsymbol{c}} \: J$ and $\nabla_{\boldsymbol{c}} \: \boldsymbol{g}$ gives rise to mathematical hurdles. Thus, we restate \eqref{eq:optproblem1} as
\begin{alignat}{3}\label{eq:optproblem2}
    \min_{\boldsymbol{x}, \boldsymbol{\theta}} \quad &J \left( \boldsymbol{u}(\boldsymbol{x}, \boldsymbol{\theta}), \boldsymbol{x}, \boldsymbol{\theta} \right)  \\
    \text{subject to} \quad &\mathbf{K}(\boldsymbol{x}, \boldsymbol{\theta}) \: \boldsymbol{u}(\boldsymbol{x}, \boldsymbol{\theta}) = \boldsymbol{f}(\boldsymbol{x}, \boldsymbol{\theta}), \nonumber \\
    &\boldsymbol{g}\left( \boldsymbol{u}(\boldsymbol{x}, \boldsymbol{\theta}), \boldsymbol{x}, \boldsymbol{\theta} \right) \leq 0, \nonumber \\
    &lb_i \leq x_i \leq ub_i, \quad \quad &&i=1,\cdots, n_x, \nonumber
\end{alignat}
where $\boldsymbol{\theta} := [ \theta_{1,1}, \cdots, \theta_{1,N_1}, \theta_{2,1}, \cdots, \theta_{2,N_2}, \cdots, \theta_{n_c,1}, \cdots, \theta_{n_c,N_{n_c}}]^T$ denotes the vector of unnormalized log-probabilities of the categorical variables formulated in \eqref{eq:theta}. The first and second indices in $\theta_{i,j}$ refer, respectively, to the variable index and its available choices' index. If for a categorical variable $i$, the probability of using its $N_i$ choices in the structure is defined by $\boldsymbol{p}_i := [p_{i,1}, \cdots, p_{i,N_i}]^T$, then 
\begin{equation}\label{eq:theta}
    \begin{aligned}
    \theta_{i,j} = \text{logit} (p_{i,j}) := \ln \left( \frac{p_{i,j}}{1 - p_{i,j}} \right), \quad j=1,\cdots, N_i,
    \end{aligned}
\end{equation}
where $p_{i,j}$ is the likelihood of choice $j$ for categorical variable $i$. Here, we opt for using unnormalized log-probabilities $\theta_{i,j}, \: i=1,\cdots, n_c, \: j=1,\cdots, N_i$, as the design variables instead of their corresponding probabilities $p_{i,j}$. Once the optimal values of $\theta_{i,j}$'s are found, their corresponding optimal probabilities $p_{i,j}$ can be computed through the Softmax function as
\begin{equation}\label{eq:thetatop}
    \begin{aligned}
    p_{i,j} = \left( \text{softmax} (\boldsymbol{\theta}_i) \right)_j := \frac{\exp(\theta_{i,j})}{\sum_{k=1}^{N_i} \exp(\theta_{i,k})}, \quad i=1,\cdots, n_c, \: j=1,\cdots, N_i.
    \end{aligned}
\end{equation}

The benefits of using $\theta_{i,j}$s are twofold. Firstly, following \eqref{eq:theta}, as opposed to $p_{i,j}$s that are between 0 and 1, unnormalized log-probabilities $\theta_{i,j}$s are unbounded continuous variables and can take any value between $-\infty$ and $+\infty$. Secondly, for each categorical variable $i$ the probabilities must satisfy $\sum_{j=1}^{N_i} p_{i,j} = 1$ at each optimization iteration, whereas working with its associated $\theta_{i,j}$s and eventually converting them to $p_{i,j}$s using \eqref{eq:thetatop} would naturally satisfy this constraint. This reduces the number of optimization constraints significantly, especially in large-scale problems.

In order to utilize gradient-based optimizers, one needs to compute the sensitivities (gradients) of objective and constraint functions with respect to $\boldsymbol{\theta}$ and $\boldsymbol{x}$ (i.e., $\nabla_{\boldsymbol{x}} J$, $\nabla_{\boldsymbol{x}} \boldsymbol{g}$, $\nabla_{\boldsymbol{\theta}} J$ and $\nabla_{\boldsymbol{\theta}} \boldsymbol{g}$). Most often, in structural engineering problems, $J$ and $\boldsymbol{g}$ are not explicit functions of the probabilities $\boldsymbol{\theta}$. For instance, for a categorical variable corresponding to a beam's material, $J$ and $\boldsymbol{g}$ are functions of the properties of the selected material such as its Young's modulus or density, not the material itself as an object. Therefore, for all categorical variables, a sampling process must be performed independently to draw a sample from each of their existing choices based on their probabilities and then compute the values of $J$ and $\boldsymbol{g}$ using that sample. The sampling process, however, in general breaks down the differentiability of these functions. This issue can be resolved by incorporating the GSM method. In the subsequent section, the GSM method and its application in having a differentiable sampling procedure are introduced. 

\section{Differentiable sampling via the GSM method}
\label{sec:gumbel}
For conciseness, the subject of the current section is presented only for $J$. The same idea can be directly applied to the constraint functions $\boldsymbol{g}$ in a similar fashion. The goal is to compute the value of $J$ and its gradient with respect to the vector of unnormalized log-probabilities $\boldsymbol{\theta}$. For the sake of brevity, we assume $\boldsymbol{\theta}$ is associated with just one categorical variable with $N$ choices and $J$ is only a function of $\boldsymbol{\theta}$ (and not $\boldsymbol{x}$ and $\boldsymbol{u}$ as in the previous section). More general cases are provided in Section \ref{sec:method}. Note that $J$ is not an explicit function of $\boldsymbol{\theta}$ but rather the continuous attributes (properties) of a sample (i.e., choice) drawn from the categorical distribution function characterized by $\boldsymbol{\theta}$. In other words, once a sample is picked from the categorical distribution of $\boldsymbol{\theta}$, the value of $J$ using the properties of this sample is computed. 

A categorical distribution function is a discrete probability distribution specifying the likelihood of the choices (classes) for a categorical random variable \cite{rohatgi2015introduction}. For instance, applying \eqref{eq:thetatop}, the categorical distribution of $\boldsymbol{\theta}$ is $p_{\boldsymbol{\theta}} := \{ ( \text{softmax} (\boldsymbol{\theta}) )_1, \cdots, ( \text{softmax} (\boldsymbol{\theta}) )_N  \}$. A \emph{one-hot} sample vector $\boldsymbol{s}$ can be generated from $p_{\boldsymbol{\theta}}$ utilizing Algorithm \ref{alg:nondiffsample}. The size of $\boldsymbol{s}$ is $N$ and its entries are all 0 except for the index corresponding to the selected class at which its entry is 1. Suppose the attributes of the selected class are collected in a vector $\boldsymbol{a}$. Since $J$ is a function of $\boldsymbol{\theta}$ through the drawn sample and its attributes, it can be formulated as $J\left(\boldsymbol{a} \left(\boldsymbol{s}(\boldsymbol{\theta}) \right) \right)$. The sensitivity of $J$ with respect to $\boldsymbol{\theta}$ therefore can be written as
\begin{equation}\label{eq:derphi}
    \begin{aligned}
    \nabla_{\boldsymbol{\theta}} J =  \left( \nabla_{\boldsymbol{\theta}} \boldsymbol{a} \right)^T \nabla_{\boldsymbol{a}} J=  \left( \nabla_{\boldsymbol{s}} \boldsymbol{a} \: \: \nabla_{\boldsymbol{\theta}} \boldsymbol{s} \right)^T \nabla_{\boldsymbol{a}} J.
    \end{aligned}
\end{equation}

The sample vector $\boldsymbol{s}$, generated by incorporating the \emph{common} sampling process described in Algorithm \ref{alg:nondiffsample}, is not differentiable with respect to $\boldsymbol{\theta}$ creating issues in computing $\nabla_{\boldsymbol{\theta}} \boldsymbol{s}$ and consequently $\nabla_{\boldsymbol{\theta}} J$ in \eqref{eq:derphi}. The step causing nondifferentiability is Step \ref{ln:index}. The GSM method resolves this issue by a simple yet remarkably efficient solution \cite{gumbel1954statistical, maddison2014sampling}. 

\begin{algorithm}[tb]
\caption{The common, nondifferentiable way of drawing samples from a categorical distribution}\label{alg:nondiffsample}
\SetAlgoLined
\KwInput{Unnormalized log-probabilities $\boldsymbol{\theta} \in \mathbb{R}^N$ associated with a categorical variable.}
\KwOutput{One-hot sample vector $\boldsymbol{s}$.}
    Initialize $\boldsymbol{s} \in \mathbb{R}^N$ by zeros\; 
    Compute the probabilities $p_i, \: i =1,\cdots,N,$ using \eqref{eq:thetatop} and put them into a (not necessarily ordered) set $p_{\boldsymbol{\theta}} := \{ p_1, \cdots, p_N  \}$\;
    Calculate the corresponding cumulative distribution function as 
    \begin{equation*}
        \overline{p}_{\boldsymbol{\theta}} := \{ 0, p_1, p_1 + p_2, \cdots, \sum_{i=1}^{N-1} p_i \};
    \end{equation*}\\
    Choose a uniformly distributed random real number $r \in [0,1]$\; \label{ln:uni}
    Find the largest index $i$ in $\overline{p}_{\boldsymbol{\theta}}$ such that $(\overline{p}_{\boldsymbol{\theta}})_i \leq r$\; \label{ln:index}
    Set the entry in $\boldsymbol{s}$ corresponding to index $i$ to 1\;
\end{algorithm}

\subsection{The GSM method}
\label{subsec:gumbeldiff}
We first introduce the GM method from which the GSM method is derived (recall that GM stands for Gumbel-Max). Define $\mathcal{I}(i)$ as an operator that returns a one-hot vector of size $N$ with an entry of 1 at index $i$ and zero elsewhere. In the GM method, a sample vector $\overline{\boldsymbol{s}}$ is generated by
\begin{equation}\label{eq:gsm}
    \overline{\boldsymbol{s}} = \mathcal{I} \left(\argmax_{i\in \{1, \cdots, N \} } \left(\theta_i + G^{(i)} \right) \right), 
\end{equation}
where $\theta_i, \: i = 1, \cdots, N,$ are unnormalized log-probabilities of the classes and $G^{(i)}, \: i = 1, \cdots, N,$ are independent and identically distributed samples (noises) generated from the standard Gumbel distribution $\text{Gumbel}(0,1)$. For each class index $i$, the noise $G^{(i)}$  can be obtained by independently sampling $\text{Gumbel}(0,1)$ through drawing a uniformly distributed random real number $r$ and setting $G^{(i)} = -\ln\left(-\ln(r) \right)$ \cite{jang2016categorical}. Therefore, instead of employing Algorithm \ref{alg:nondiffsample}, which is the common, nondifferentiable way of drawing samples from a categorical distribution, $\overline{\boldsymbol{s}}$ can be drawn from $\boldsymbol{\theta}$ by perturbing each entry of $\boldsymbol{\theta}$ via independently adding a Gumbel noise to them, selecting the index corresponding to the largest perturbed entry of $\boldsymbol{\theta}$, and returning the sample associated with that index. The sample vector $\overline{\boldsymbol{s}}$ produced this way has theoretically the same distribution as the categorical distribution associated with $\boldsymbol{\theta}$ (i.e., $\overline{\boldsymbol{s}} \sim p_{\boldsymbol{\theta}}$). For the sake of conciseness, the proof of this statement is provided in Appendix \ref{appen:gm}. It is of the utmost importance to note that in \eqref{eq:gsm} only the noise generated from the standard Gumbel distribution---and not any other types of probability distributions---results in a sample vector $\overline{\boldsymbol{s}} \sim p_{\boldsymbol{\theta}}$. Interested readers may refer to \cite{gumbel1954statistical, jang2016categorical, huijben2022review} for further details about the GM method and other versions of the proof.  

Since the $\argmax$ function in \eqref{eq:gsm} is not differentiable, Jang et al. \cite{jang2016categorical} proposed a continuous, differentiable approximation (relaxation) to this function by introducing a temperature parameter $\tau$ and utilizing a so-called soft one-hot sample vector $\widetilde{\boldsymbol{s}}$ with entries 
\begin{equation}\label{eq:zvec}
    \widetilde{s}_i := \frac{\exp \left( (\theta_{i} + G^{(i)}) / \tau \right)}{\sum_{j=1}^{N} \exp \left(( \theta_{j} + G^{(j)}) / \tau \right)}, \quad i=1,\cdots, N,
\end{equation}
or put it shortly $\widetilde{\boldsymbol{s}} = \text{softmax} \left( (\boldsymbol{\theta} + \boldsymbol{G}) / \tau  \right)$ with $\boldsymbol{G} := [G^{(1)}, \cdots, G^{(N)}]^T$. Although $\widetilde{s}_i$ entries in \eqref{eq:zvec} have the advantage of being differentiable with respect to $\theta_i, \: i=1,\cdots,N$, there are two concerns with having a soft one-hot sample vector $\widetilde{\boldsymbol{s}}$ in the GSM method. Firstly, due to the applied relaxation, $\widetilde{\boldsymbol{s}}$ does not exactly follow the $p_{\boldsymbol{\theta}}$ distribution, as opposed to $\overline{\boldsymbol{s}}$ in \eqref{eq:gsm}. However, as $\tau \rightarrow 0$, this soft one-hot vector becomes a true one-hot vector and subsequently, the samples generated by the GSM method would have the exact $p_{\boldsymbol{\theta}}$ distribution \cite{jang2016categorical}. Therefore, at the beginning of the optimization routine, one can set $\tau$ to a high temperature (in a relative sense) and using an annealing scheme reduce it to a small nonzero value as the optimization progresses. For machine learning applications, the typical range suggested for $\tau$ is between 100 to 0.1 (e.g., \cite{baevski2019vq, yang2019modeling, guo2020learning, kang2020operation}). More details about the annealing scheme and temperature range used in this study are provided in Section \ref{sec:case}.

Secondly, with regard to the concerns facing the GSM method, having a soft one-hot vector $\widetilde{\boldsymbol{s}}$ leads to generating a sample vector that is a combination of all choices in the target categorical variable which oftentimes does not bear any physical realization, causing problems in computing the value of $J$ and its gradients. For example, for the applications considered in this paper, we cannot have a beam cross-sectional profile that is a combination of I- and U-profiles. In such scenarios, the so-called \emph{straight-through} GSM \cite{jang2016categorical} can be adopted to generate a true one-hot sample vector 
\begin{equation}\label{eq:yvec}
    \widehat{\boldsymbol{s}} := \mathcal{I} \left(\argmax_{i\in \{1, \cdots, N \} } (\widetilde{s}_i) \right),
\end{equation}
and to compute the value of $J$ using this vector. Then, for the sensitivity analysis in \eqref{eq:derphi}, since $\widehat{\boldsymbol{s}}$ is not differentiable with respect to $\boldsymbol{\theta}$, $\nabla_{\boldsymbol{\theta}} \widetilde{\boldsymbol{s}}$ instead of $\nabla_{\boldsymbol{\theta}} \widehat{\boldsymbol{s}}$ is utilized. As the optimization advances and $\widetilde{\boldsymbol{s}}$ approaches toward being a true one-hot vector, the discrepancy between $\nabla_{\boldsymbol{\theta}} \widetilde{\boldsymbol{s}}$ and $\nabla_{\boldsymbol{\theta}} \widehat{\boldsymbol{s}}$ diminishes. Ultimately, following \eqref{eq:derphi} and generating $\widetilde{\boldsymbol{s}}$ using the GSM method, the gradient of $J$ with respect to $\theta_j, \: =1,\cdots, N$, reads
\begin{equation}\label{eq:derphi2}
    \begin{aligned}
    \nabla_{\theta_j} J &=  \nabla_{\theta_j} \boldsymbol{a} \cdot \nabla_{\boldsymbol{a}} J=  \left( \nabla_{\widetilde{\boldsymbol{s}}} \boldsymbol{a} \: \: \nabla_{\theta_j} \widetilde{\boldsymbol{s}} \right) \cdot \nabla_{\boldsymbol{a}} J = \nabla_{\boldsymbol{a}} J \cdot \sum_{i=1}^N \frac{\partial \boldsymbol{a}}{\partial \widetilde{s}_i} \frac{\partial \widetilde{s}_i}{\partial \theta_j}, \quad j=1,\cdots, N,
    \end{aligned}
\end{equation}
where 
\begin{equation}\label{eq:dz}
    \begin{aligned}
    \frac{\partial \widetilde{s}_i}{\partial \theta_j} = \begin{cases}
			\left(1 - \widetilde{s}_i \right) \widetilde{s}_i, & \text{if $i=j$}\\
            -\widetilde{s}_i \widetilde{s}_j, & \text{otherwise}
		 \end{cases},
    \end{aligned}
\end{equation}
is easily derived using \eqref{eq:zvec}. Note that although $G^{(i)}, \: i =1,\cdots, N,$ which are random Gumbel noises appear in computing $\widetilde{s}_i$ values in \eqref{eq:zvec}, they do not cause any differentiability issues in \eqref{eq:derphi2} and \eqref{eq:dz}. Assuming $\nabla_{\boldsymbol{a}} J$ and $\nabla_{\widetilde{\boldsymbol{s}}} \boldsymbol{a}$ are known, Algorithm \ref{alg:diffsample} encapsulates the steps involved in the GSM method for generating sample vectors $\widetilde{\boldsymbol{s}}$ and $\widehat{\boldsymbol{s}}$, as well as computing the value of a function $J$ that through $\widetilde{\boldsymbol{s}}$ is implicitly a function of unnormalized log-probabilities $\boldsymbol{\theta}$ corresponding to a categorical distribution and eventually calculating $\nabla_{\boldsymbol{\theta}} J$. The details of computing $\nabla_{\boldsymbol{a}} J$ and $\nabla_{\widetilde{\boldsymbol{s}}} \boldsymbol{a}$ are laid out in the next section.

\begin{algorithm}[!t]
\caption{The GSM method for computing the value and gradient of a function of a categorical distribution}\label{alg:diffsample}
\SetAlgoLined
\KwInput{Function $J$ and its gradients $\nabla_{\boldsymbol{a}} J$ and $\nabla_{\widetilde{\boldsymbol{s}}} \boldsymbol{a}$, unnormalized log-probabilities $\boldsymbol{\theta} \in \mathbb{R}^N$ associated with a categorical variable, temperature parameter $\tau$.}
\KwOutput{$J$ and $\nabla_{\boldsymbol{\theta}} \: J$.}
    Generate samples $G^{(i)}, \: i=1,\cdots, N,$ from the standard Gumbel distribution $\text{Gumbel}(0,1)$\;
    Compute $\widetilde{s}_i, \: i=1,\cdots, N,$ through \eqref{eq:zvec}\;
    Calculate $\widehat{\boldsymbol{s}}$ using \eqref{eq:yvec}\;
    Get $J$ using $\widehat{\boldsymbol{s}}$\;
    Compute $\partial \widetilde{s}_i / \partial \theta_j, \: i,j=1,\cdots, N$, using \eqref{eq:dz}\;
    Get $\nabla_{\boldsymbol{\theta}} \: J$ via \eqref{eq:derphi2}\;
\end{algorithm}

\section{Optimization methodology}
\label{sec:method}

In Section \ref{sec:gumbel}, the GSM method was described in its most general form regardless of the application. Henceforth, to establish a more oriented dialogue, we limit the scope to structural problems and optimization functions arising in such applications. Considering the optimization problem stated in \eqref{eq:optproblem2}, in this section, we develop a scheme to compute the sensitivities of the optimization objective and constraint functions ($J$ and $\boldsymbol{g}$, respectively) with respect to both $\boldsymbol{x}$ and $\boldsymbol{\theta}$. Without loss of generality, we only focus on computing $\nabla_{\boldsymbol{x}} J$ and $\nabla_{\boldsymbol{\theta}} J$. Computing $\nabla_{\boldsymbol{x}} \boldsymbol{g}$ and $\nabla_{\boldsymbol{\theta}} \boldsymbol{g}$ proceeds similarly. 

For the structural problems considered in this article, continuous design variables $\boldsymbol{x}$ could be components' cross-sectional dimensions (or properties), their orientation, length and material properties. The categorical variables may be material and cross-sectional profile choices of each component. The categorical variables are determined by their corresponding continuous attributes. For example, material choices are characterized through their Young's modulus, Poisson's ratio and density. The objective and constraint functions may be the structure's mass, its compliance, nodal displacement and maximum stress to name a few. The proposed optimization framework and required derivations are elucidated in a general format such that other types of design variables and functions can be incorporated in a straightforward manner. 

For the sake of brevity, suppose the aim is to compute the sensitivities of $J$ with respect to a single continuous variable $x_l, \: l = 1, \cdots, n_x$ and unnormalized log-probabilities $\boldsymbol{\theta}_m := [\theta_{m,1}, \cdots, \theta_{m,N_m}]^T$ associated with a single categorical variable $c_m, \: m=1,\cdots, n_c$. Subsections \ref{subsec:sensitivitycon} and \ref{subsec:sensitivitycat} present the procedures for computing $\nabla_{x_l} J$ and $\nabla_{\boldsymbol{\theta}_m} J$, respectively. Once sensitivities of $J$ are calculated, they can be utilized for optimizing the frame structures using the optimization approach proposed in Subsection \ref{subsec:scheme}.

\subsection{Sensitivity analysis for continuous variables}
\label{subsec:sensitivitycon}
Recall from Section \ref{sec:problemdef} that $J$ can explicitly and implicitly be a function of $x_l$ (and $\boldsymbol{\theta}_m$) through $\boldsymbol{u}$. Accordingly, the gradient of $J$ with respect to $x_l$ is expressed as
\begin{equation}
    \begin{aligned}\label{eq:sensjx}
    \nabla_{x_l} J &= \frac{\partial J}{\partial x_l} + \nabla_{x_l} \boldsymbol{u} \cdot \nabla_{\boldsymbol{u}} J, \quad l = 1, \cdots, n_x, \\
\end{aligned}
\end{equation}
where $\partial J / \partial x_l$ and $\nabla_{\boldsymbol{u}} J$ which are, respectively, explicit derivatives of $J$ with respect to $x_l$ and $\boldsymbol{u}$ can be carried out in a simple manner knowing the function formulation, thus not further elaborated in this paper. For example, if $J = 0.5 \boldsymbol{u}^T \mathbf{K} \boldsymbol{u}$, then $\partial J / \partial x_l = 0.5 \boldsymbol{u}^T \left( \nabla_{x_l}  \mathbf{K} \right) \boldsymbol{u}$ and $\nabla_{\boldsymbol{u}} J = \mathbf{K} \boldsymbol{u}$. The onerous term in \eqref{eq:sensjx} is $\nabla_{\boldsymbol{u}} J$ for which we appeal to the \emph{adjoint} method \cite{ebrahimi2019design, cao2003adjoint} well-known to be advantageous over the direct differentiation method for optimizing problems with several design variables. The process is presented for a general $J$ that may be a linear or nonlinear function of its arguments. Using the discretized governing equation $\mathbf{K} \boldsymbol{u} - \boldsymbol{f} = \boldsymbol{0}$ and differentiating it with respect to $x_l$ yields
\begin{equation}\label{eq:govder}
    \mathbf{K} \: \nabla_{x_l} \boldsymbol{u} + \left( \nabla_{x_l} \mathbf{K} \right) \boldsymbol{u} - \nabla_{x_l} \boldsymbol{f} = \boldsymbol{0},
\end{equation}
where it is assumed that the external force vector $\boldsymbol{f}$ is only a function of $x_l$ not $\boldsymbol{u}$. Since the left-hand side of \eqref{eq:govder} is zero, multiplying it by any arbitrary vector of a conforming size to the governing equation also results in zero. Denote this multiplier vector, called the adjoint variable vector, by $\boldsymbol{\lambda}_J \in \mathbb{R}^{n_u}$. Applying it to \eqref{eq:govder} and employing the result in \eqref{eq:sensjx} leads to
\begin{equation}\label{eq:sensjxgov}
    \begin{aligned}
    \nabla_{x_l} J &= \frac{\partial J}{\partial x_l} + \nabla_{x_l} \boldsymbol{u} \cdot \nabla_{\boldsymbol{u}} J - \boldsymbol{\lambda}_J \cdot \left( \mathbf{K} \: \nabla_{x_l} \boldsymbol{u} + \left( \nabla_{x_l} \mathbf{K} \right) \boldsymbol{u} - \nabla_{x_l} \boldsymbol{f} \right), \quad l = 1, \cdots, n_x.
    \end{aligned}
\end{equation}
This equation holds regardless of the value of $\boldsymbol{\lambda}_J$; ergo, it can be chosen such that $\nabla_{x_l} \boldsymbol{u}$ in \eqref{eq:sensjxgov} vanishes. Meaning
\begin{equation}\label{eq:adjoint}
    \mathbf{K} \boldsymbol{\lambda}_J = \nabla_{\boldsymbol{u}} J.
\end{equation}
Once $\boldsymbol{\lambda}_J$ is found through \eqref{eq:adjoint}, putting it in \eqref{eq:sensjxgov} gives
\begin{equation}\label{eq:sensjxfinal}
    \nabla_{x_l} J = \frac{\partial J}{\partial x_l} - \boldsymbol{\lambda}_J \cdot \left( \left( \nabla_{x_l} \mathbf{K} \right) \boldsymbol{u} - \nabla_{x_l} \boldsymbol{f} \right), \quad l = 1, \cdots, n_x.
\end{equation}
This is the final equation for computing the sensitivity of $J$ with respect to continuous design variables. Note that solving \eqref{eq:adjoint} can be carried out quite efficiently as factorizing $\mathbf{K}$, the arduous step in the solution process, has already been performed in solving the governing equation $\mathbf{K} \boldsymbol{u} = \boldsymbol{f}$. Furthermore, there is no dependence on $x_l$ or any other continuous design variable in \eqref{eq:adjoint}; hence \eqref{eq:adjoint} is solved only once (and not for each variable individually), justifying the application of the adjoint method for problems with several design variables. The adjoint equation for each constraint function $g_i, \: i=1,\cdots,n_g$, is formulated as
\begin{equation}\label{eq:adjointg}
    \mathbf{K} \boldsymbol{\lambda}_{g_i} = \nabla_{\boldsymbol{u}} g_i.
\end{equation}

\subsection{Sensitivity analysis for categorical variables}
\label{subsec:sensitivitycat}
With respect to $\boldsymbol{\theta}_m$, the gradient of $J$ using \eqref{eq:derphi} reads
\begin{equation}
    \begin{aligned}\label{eq:sensjtheta}
    \nabla_{\boldsymbol{\theta}_m} J &= \left( \nabla_{\widetilde{\boldsymbol{s}}_m} \boldsymbol{a} \: \: \nabla_{\boldsymbol{\theta}_m} \widetilde{\boldsymbol{s}}_m \right)^T \nabla_{\boldsymbol{a}} J = \nabla_{\boldsymbol{a}} J \cdot \left( \nabla_{\widetilde{\boldsymbol{s}}_m} \boldsymbol{a} \: \: \nabla_{\boldsymbol{\theta}_m} \widetilde{\boldsymbol{s}}_m \right), \quad m = 1, \cdots, n_c,
    \end{aligned}
\end{equation}
where $\widetilde{\boldsymbol{s}}_m$ is the sample vector drawn from the categorical probability distribution characterized by $\boldsymbol{\theta}_m$ employing the GSM method. Calculating $\nabla_{\boldsymbol{\theta}_m} \widetilde{\boldsymbol{s}}_m$ can be performed incorporating the procedure developed in Section \ref{sec:gumbel} and \eqref{eq:dz}. We first focus on handling $\nabla_{\widetilde{\boldsymbol{s}}_m} \boldsymbol{a}$ and then $\nabla_{\boldsymbol{a}} J$ in \eqref{eq:sensjtheta}. Naturally, each entry of $\widetilde{\boldsymbol{s}}_m$ corresponds to a class which can be characterized by some continuous attributes (properties). If $\widetilde{\boldsymbol{s}}_m$ is associated with an isotropic material its entries are determined by Young's modulus and Poisson's ratio. On the other hand, if $\widetilde{\boldsymbol{s}}_m$ is linked to a cross-sectional profile, the relevant continuous attributes are cross-sectional properties such as area and second moments of area. It is important to note that all the choices considered for a categorical variable must be characterized using the same set of attributes. 

Suppose $\widetilde{\boldsymbol{s}}_m$ corresponds to a categorical variable with $N_m$ choices. Let $\boldsymbol{a}_k := [a_{1,k}, \cdots, a_{n_m,k}]^T, \: k=1,\cdots, N_m$, be the vector of size $n_m$ containing the continuous attributes of each choice. Also, define $\mathbf{A}_m \in \mathbb{R}^{n_m \times N_m}$ the attribute matrix of this categorical variable as
\begin{equation}\label{eq:attmat}
    \mathbf{A}_m := \begin{bmatrix} a_{1,1} & \cdots & a_{1,N_m} \\
    \vdots & \ddots & \vdots \\
    a_{n_m, 1} & \cdots & a_{n_m, N_m} \end{bmatrix}.
\end{equation}
If $\widetilde{\boldsymbol{s}}_m$ is a hard one-hot vector, the attributes of the selected class in this categorical variable can be picked out by 
\begin{equation}\label{eq:att}
    \boldsymbol{a} = \mathbf{A}_m \widetilde{\boldsymbol{s}}_m.
\end{equation}
As mentioned earlier in Section \ref{sec:gumbel}, for running the structural simulation and computing the optimization functions, $\widehat{\boldsymbol{s}}_m$ (given by \eqref{eq:yvec}) instead of $\widetilde{\boldsymbol{s}}_m$ needs to be used in \eqref{eq:att}. During the sensitivity analysis, however, $\widetilde{\boldsymbol{s}}_m$ itself is employed leading to $\nabla_{\widetilde{\boldsymbol{s}}_m} \boldsymbol{a} = \mathbf{A}_m$. As the optimization advances, $\widetilde{\boldsymbol{s}}_m$ approaches $\widehat{\boldsymbol{s}}_m$. Finally, inserting this relation into \eqref{eq:sensjtheta} yields
\begin{equation}\label{eq:djdsfinal}
    \begin{aligned}
    \nabla_{\boldsymbol{\theta}_m} J &= \nabla_{\boldsymbol{a}} J \cdot \left( \nabla_{\widetilde{\boldsymbol{s}}_m} \boldsymbol{a} \: \: \nabla_{\boldsymbol{\theta}_m} \widetilde{\boldsymbol{s}}_m \right) = \nabla_{\boldsymbol{a}} J \cdot \left( \mathbf{A}_m \:\: \nabla_{\boldsymbol{\theta}_m} \widetilde{\boldsymbol{s}}_m \right), \quad m = 1, \cdots, n_c. \\
    \end{aligned}
\end{equation}

In this equation, $\nabla_{\boldsymbol{a}} J$ is the sensitivity of $J$ with respect to the $\boldsymbol{a}$ associated with the continuous attributes of the class with the highest unnormalized log-probability at a given optimization iteration (i.e., the attributes of the sample generated by the straight-through GSM). Since $\boldsymbol{a}$ is a continuous variable, computing $\nabla_{\boldsymbol{a}} J$ can be carried out by adopting the adjoint method described in the earlier subsection. In other words,
\begin{equation}\label{eq:djdai}
    \nabla_{a_i} J = \frac{\partial J}{\partial a_i} - \boldsymbol{\lambda}_J^T \left( \left( \nabla_{a_i} \mathbf{K} \right) \boldsymbol{u} - \nabla_{a_i} \boldsymbol{f} \right), \quad i=1,\cdots, n_m,
\end{equation}
with the same $\boldsymbol{\lambda}_J$ found by \eqref{eq:adjoint}. Equation \eqref{eq:djdsfinal} is the final equation for computing the sensitivity of $J$ with respect to $\boldsymbol{\theta}_m$. To further clarify the presented procedure for computing $\nabla_{\boldsymbol{\theta}_m} J$, we provide the following example. Suppose an optimization problem whose categorical design variables are the cross-sectional profiles of its beams. Assume for one of the variables the choices are circular and rectangular profiles each characterized by their area $A$, second moments of area $I_{yy}$ and $I_{zz}$ and torsion constant $K$. Also, let $\boldsymbol{\theta} = [\theta_1, \theta_2]^T$ be the unnormalized log-probabilities associated with each profile at a given optimization iteration. For this variable, the attribute matrix $\mathbf{A}$ reads
\begin{equation}
    \mathbf{A} := \begin{bmatrix} 
    A_{\text{circle}} & A_{\text{rectangle}} \\ 
    I_{yy_\text{circle}} & I_{yy_\text{rectangle}} \\ 
    I_{zz_\text{circle}} & I_{zz_\text{rectangle}} \\ 
    K_{\text{circle}} & K_{\text{rectangle}}
    \end{bmatrix}.
\end{equation}

There may be two scenarios for $\theta_1$ and $\theta_2$ at any optimization iteration:
\begin{enumerate}
    \item $\theta_1 \geq \theta_2$: leading to $\widetilde{s}_1 \geq \widetilde{s}_2$ using \eqref{eq:zvec} and $\widehat{\boldsymbol{s}} = [1, 0]^T$ according to \eqref{eq:yvec}.
    \item $\theta_1 < \theta_2$: resulting in $\widetilde{s}_1 < \widetilde{s}_2$ based on \eqref{eq:zvec} and $\widehat{\boldsymbol{s}} = [0, 1]^T$ utilizing \eqref{eq:yvec}.
\end{enumerate}
In the former situation, $J$ and $\nabla_{\boldsymbol{a}} J$ in \eqref{eq:djdsfinal} are computed employing the circular profile and its continuous attributes, otherwise, the rectangular profile is adopted. Finally, $\nabla_{\boldsymbol{\theta}} J$ is obtained through $\nabla_{\widetilde{\boldsymbol{s}}_m} \boldsymbol{a} = \mathbf{A}$, finding $\nabla_{\boldsymbol{a}} J$ via \eqref{eq:djdai} and calculating $\nabla_{\boldsymbol{\theta}} \widetilde{\boldsymbol{s}}$ applying the GSM method. We once again highlight that as the optimization moves toward convergence, $\widetilde{\boldsymbol{s}}$ approaches a true one-hot vector and the errors due to the applied relaxations inherent in the GSM method dwindle.

\subsection{Optimization scheme}
\label{subsec:scheme}
The proposed optimization routine---named GSMO standing for Gumbel-Softmax optimization---for frame (or truss) structures with mixed categorical and continuous design variables is presented in Algorithm \ref{alg:optscheme}. Although this scheme is described for a particular class of structures undergoing linear elasticity behavior, it can be applied with minor modifications to other problems (not necessarily structural) with mixed categorical and continuous design variables governed by a different set of physics equations. 

\begin{algorithm}[!htb]
\caption{The GSMO scheme for optimizing frame/truss structures with mixed categorical and continuous design variables}\label{alg:optscheme}
\SetAlgoLined
\KwInput{Objective function $J$, constraint functions $\boldsymbol{g}$, continuous design variables $\boldsymbol{x}$ and their bounds, categorical design variables $\boldsymbol{c}$ and their available choices, GSM annealing scheme.}
\KwOutput{Optimum design.}
    Initialize $\boldsymbol{x}$\;
    Initialize $\boldsymbol{\theta}_i, \: i=1,\cdots,n_c,$ for each categorical variable $c_i$ by entry values of zero assuming equal probabilities for the choices in each $c_i$\;
    \While{not converged}
    {
        \For{$i=1, \cdots, n_c$}{ 
            Generate Gumbel noises $G^{(j)}, \:j=1,\cdots,N_i,$ from the standard Gumbel distribution $\text{Gumbel}(0,1)$\;
            Compute $(\widetilde{\boldsymbol{s}}_{i})_j, \:j=1,\cdots,N_i,$ using \eqref{eq:zvec} and $\nabla_{\boldsymbol{\theta}_i} \widetilde{\boldsymbol{s}}_i$ via \eqref{eq:dz}\;\label{st:dsdtheta}
            Calculate $\widehat{\boldsymbol{s}}_i$ through \eqref{eq:yvec}\;
        }
        Solve the governing equations $\mathbf{K} \boldsymbol{u} = \boldsymbol{f}$ using $\widehat{\boldsymbol{s}}_i, \: i=1, \cdots, n_c,$\ and get $\boldsymbol{u}$\;
        Compute $J$ and $\boldsymbol{g}$ values using $\widehat{\boldsymbol{s}}_i, \: i=1, \cdots, n_c,$\ and $\boldsymbol{u}$\;
        Calculate $\nabla_{\boldsymbol{u}} J$ then find $\boldsymbol{\lambda}_J$ by solving \eqref{eq:adjoint}\;\label{st:adjj}
        Compute $\nabla_{\boldsymbol{u}} g_i, \: i=1,\cdots, n_g,$ then find $\boldsymbol{\lambda}_{g_i}, \: i=1,\cdots, n_g,$ by solving \eqref{eq:adjointg}\;\label{st:adjg}
        Calculate $\partial J / \partial x_i$, $\partial \boldsymbol{g} / \partial x_i$, $\partial \mathbf{K} / \partial x_i$ and $\partial \boldsymbol{f} / \partial x_i, \: i=1,\cdots,n_x$\;
        Get $\nabla_{\boldsymbol{x}} J$ and $\nabla_{\boldsymbol{x}} \boldsymbol{g}$ incorporating the corresponding adjoint vectors and \eqref{eq:sensjxfinal}\;\label{st:senx}
        \For{$i=1, \cdots, n_c$}{
            Form the attribute matrix $\mathbf{A}_i$ as in \eqref{eq:attmat}\;
            Calculate $\partial J / \partial \boldsymbol{a}_i$, $\partial \boldsymbol{g} / \partial \boldsymbol{a}_i$, $\partial \mathbf{K} / \partial \boldsymbol{a}_i$ and $\partial \boldsymbol{f} / \partial \boldsymbol{a}_i$ using the attributes of the selected class for this categorical variable\;
            Compute $\nabla_{\boldsymbol{a}_i} J$ and $\nabla_{\boldsymbol{a}_i} \boldsymbol{g}$ utilizing the adjoint vectors found in Steps \ref{st:adjj} and \ref{st:adjg}, $\nabla_{\boldsymbol{\theta}_i} \widetilde{\boldsymbol{s}}_i$ found in Step \ref{st:dsdtheta} and employing \eqref{eq:djdai}\;
            Get $\nabla_{\boldsymbol{\theta}_i} J$ and $\nabla_{\boldsymbol{\theta}_i} \boldsymbol{g}$ through \eqref{eq:djdsfinal}\;\label{st:sentheta}
        }
     Update $\boldsymbol{x}$ and $\boldsymbol{\theta}_i, \: i=1,\cdots,n_c$, using the sensitivities found in Steps \ref{st:senx} and \ref{st:sentheta}\;
    }
\end{algorithm} 

An important fact about GSMO in Algorithm \ref{alg:optscheme} is that there is no stratification between continuous and categorical design variables. In other words, both classes of variables are handled concurrently in each optimization iteration, thus giving no priority to either of them. This is unlike the bilevel optimization routines where categorical and continuous variables are treated at different levels (stages) \cite{allaire2016stacking, islam2017multimodal, barjhoux2020bi, cheong2021optimal}. In such techniques, first, the categorical variables are taken into account and the structure is optimized updating only those. Then, in the second level, the continuous design variables are handled. A shortcoming of these approaches is that each variable type is optimized sequentially at one level at a time, which may lead to underperformance. Nevertheless, a bilevel version of GSMO, named BiGSMO, is provided in Algorithm \ref{alg:bioptscheme} of Appendix \ref{appen:bigsmo} and assayed against GSMO in the forthcoming section.

\section{Case studies}
\label{sec:case}
Prior to introducing the case studies, we note a few remarks. Since the GSM method relies on generating Gumbel samples which is a stochastic process, the optimum solution found through GSMO (and BiGSMO) may vary every time the optimization is executed. Therefore, GSMO (and BiGSMO) shares a similarity in this aspect with stochastic optimizers such as GA. Hence, to make a fair comparison with stochastic methods, we run each of the case studies 10 times and report the best, average and standard deviation of optimum solutions. As for the temperature $\tau$ in \eqref{eq:zvec} and its annealing scheme, an initial temperature of 100 scaled by 0.9 per iteration is adopted. The minimum temperature is kept at 0.01. Other options for the temperature and its annealing scheme are described in \cite{huijben2022review}. Also, a step size of $10^{-3}$ is set for the optimization routine. These choices can be seen as hyper-parameters for GSMO. Moreover, an identical initial probability value is assigned to the available choices of each categorical variable. Further remarks about them are provided in the conclusion section.

Unfortunately, the relevant literature to this article lacks a repository of well-documented benchmarks, particularly for beam structures \cite{stolpe2016truss}. Nevertheless, to study the performance of GSMO, its results are compared to those reported in \cite{shahabsafa2018novel} using different optimization techniques (which includes GA, particle swarm optimization (PSO) \cite{kaveh2009particle}, mine blast optimization (MBO) \cite{sadollah2012mine}, water cycle optimization (WCO) \cite{sadollah2015water}, colliding bodies optimization (CBO)\cite{kaveh2015comparative}, differential evolution (DE) \cite{ho2016adaptive} and neighborhood search (NS) \cite{shahabsafa2018novel}) for a 72-bar truss structure (first case study) and those produced by a GA scheme for the other case studies. For the implemented GA, the population size of 10 times the number of design variables, crossover rate of 0.9 and mutation rate of 0.1 are selected. Furthermore, the constraints are enforced using a penalty method with a penalty factor of 1000. The maximum number of iterations for both GSMO and GA is set to 100. The case studies with a mix of continuous and categorical variables are also run by BiGSMO considering 10 iterations for both outer and inner loops (Lines \ref{ln:outer} and \ref{ln:inner} in Algorithm \ref{alg:bioptscheme}). All tests are run on a single desktop computer with an Intel Core i9 Processor at 2.40GHz with 8 cores and 16 threads.

\subsection{72-bar truss structure}
\label{subsec:truss}

\begin{figure}[!ht]
\centering
\includegraphics[width=0.6\textwidth]{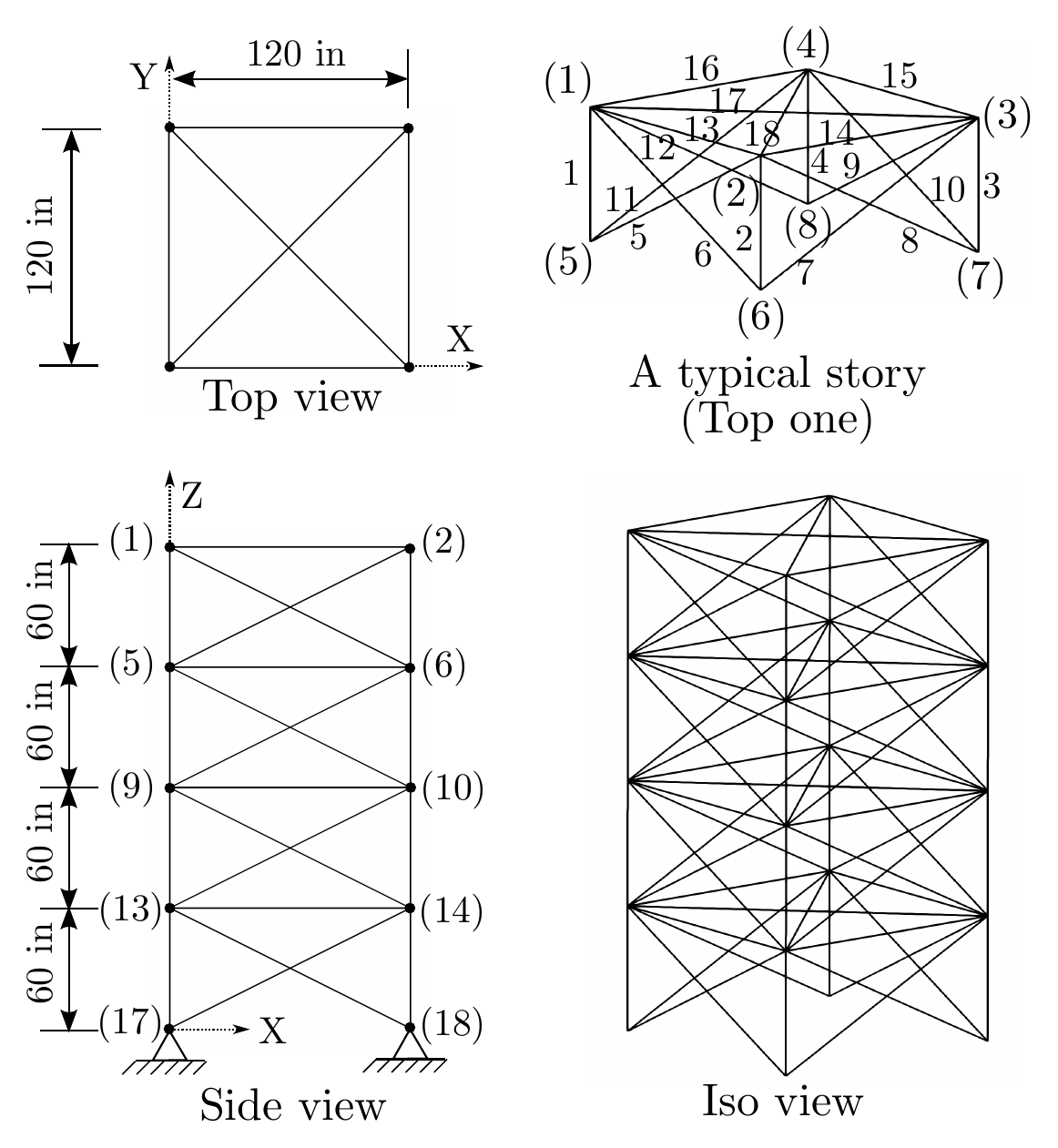}
\caption{The 72-bar truss structure. The typical story is repeated 4 times along the Z axis. The node and element numbering also follows suit.}
\label{fig:trussgeo}
\end{figure}

The 72-bar truss structure \cite{haftka2012elements, shahabsafa2018novel} depicted in Figure \ref{fig:trussgeo} undergoes two load cases. In the first one, a force of $\boldsymbol{f} = [5000, \: 5000, \: 0]^T$ lbf is applied to Node 1, and in the second load case, a force of $\boldsymbol{f} = [0, \: 0, \: -5000]^T$ lbf is exerted on Nodes 1, 2, 3 and 4. The objective function for this problem is to minimize the structure's mass and the design variables are the cross-sectional areas of all trusses (hence 72 variables in aggregate) selected from a discrete set given in Table \ref{table:trusscs}. All members are made of an aluminum alloy with a density of $0.1 \: \mathrm{lb/in^3}$, Young's modulus of $10^7$ psi and yield stress of $25000$ psi. For both load cases, the stress in all trusses must remain below the given yield stress and the displacement of Nodes 1, 2, 3 and 4 along X and Y axes must be bounded by $\pm 0.25$ in.

\begin{table}[b]
\caption{\label{table:trusscs}Categorical choices of the cross-sectional areas for the 72-bar truss structure}
\begin{center}
\begin{tabular}{l}
\toprule
\textbf{Areas ($\mathrm{in^2}$)}\\[0.2em]
\hline
0.111, 0.141, 0.196, 0.250, 0.307, 0.391, 0.442, 0.563, 0.602, 0.766, 0.785, \\[0.2em]
0.994, 1.000, 1.228, 1.266, 1.457, 1.563, 1.620, 1.800, 1.990, 2.130, 2.380, \\[0.2em]
2.620, 2.630, 2.880, 2.930, 3.090, 3.130, 3.380, 3.470, 3.550, 3.630, 3.840, \\[0.2em]
3.870, 3.880, 4.180, 4.220, 4.490, 4.590, 4.800, 4.970, 5.120, 5.740, 7.220, \\[0.2em]
7.970, 8.530, 9.300, 10.85, 11.50, 13.50, 13.90, 14.20, 15.50, 16.00, 16.90, \\[0.2em]
18.80, 19.90, 22.00, 22.90, 24.50, 26.50, 28.00, 30.00, 33.50 \\[0.2em]
\bottomrule
\end{tabular}
\end{center}
\end{table}

\begin{figure}[ht]
\centering
    \includegraphics[width=0.8\textwidth]{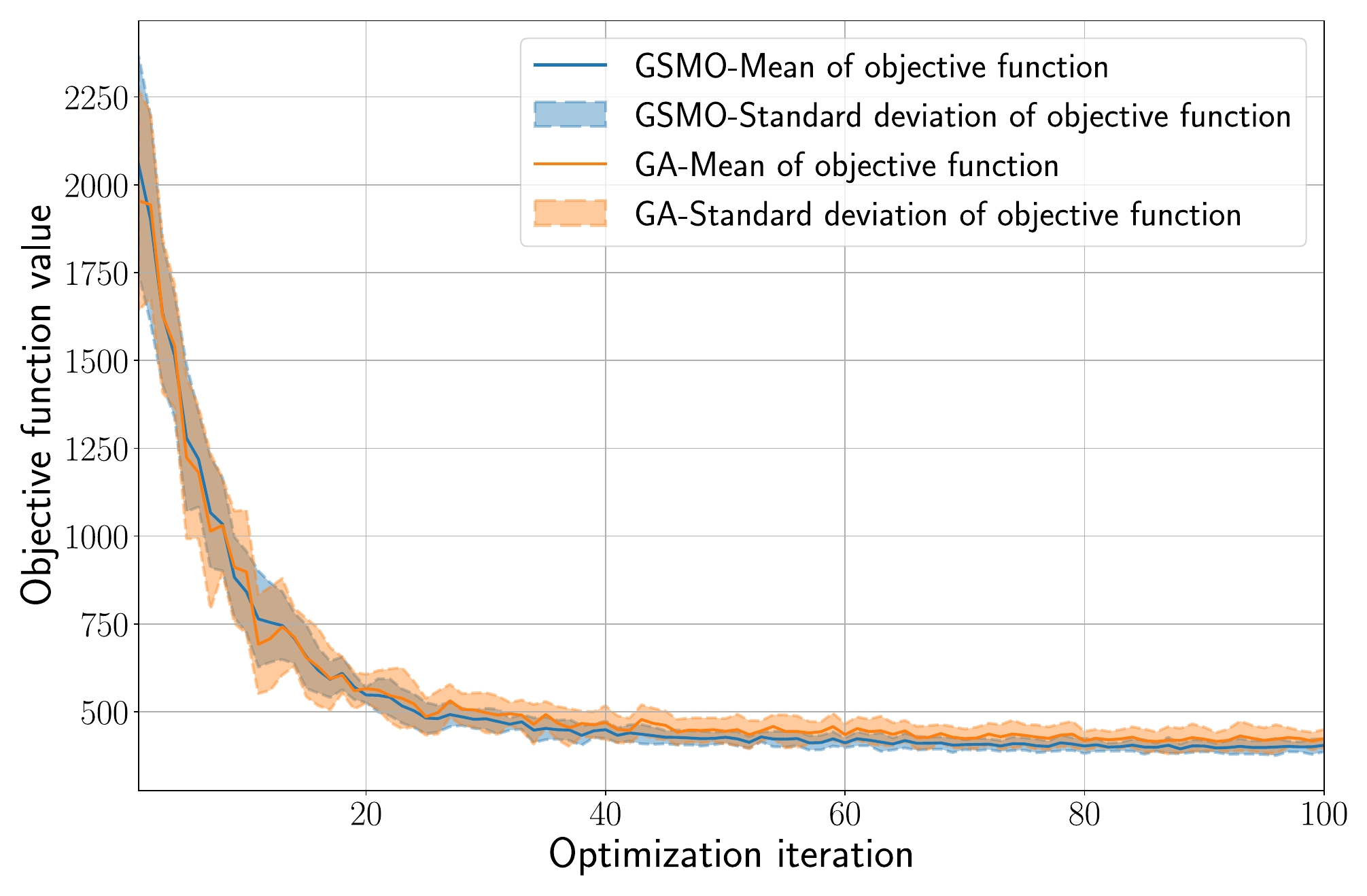}
    \caption{Convergence plot of the 10 GSMO and GA runs for the truss structure}\label{fig:trussconv}
\end{figure}

Figure \ref{fig:trussconv} shows the convergence plot of the mean and standard deviation of the 10 GSMO and GA runs. Accordingly, the optimization routine advances smoothly and reaches convergence quite rapidly using both algorithms. Table \ref{table:trussres} presents the optimum solution obtained by GSMO (the best one amongst the 10 runs), GA and those of other approaches reported in \cite{shahabsafa2018novel}. For this problem, since the design variables are only categorical, in essence, there is no difference between GSMO and BiGSMO and the result of the latter is not provided. As can be seen, the best solution generated by GSMO has a slight lead compared to those reported previously. The mean and standard deviation of the 10 optimal results by GSMO are respectively 394.31 and 11.14, which may be taken as an affirmation of the fact that although a sampling process is involved in GSMO, the method consistently finds close optimum solutions (we have reviewed this claim in the subsequent case studies as well). In this problem, perhaps the most significant advantage of GSMO is its computational time. Since GSMO is a gradient-based approach, it requires only one FE solve per optimization iteration leading to a total of 100 FE solves for each run. For this example, each GSMO run takes about 0.9 seconds on average on the computer specified earlier.

\begin{table}[ht]
\caption{\label{table:trussres}Categorical choices of the cross-sectional areas ($\mathrm{in}^2$) for the truss structure found by various optimization methods in the literature}
\begin{center}
\begin{tabular}{ccccccccc}
\toprule[0.1em]
Member ID & \begin{tabular}{@{}c} PSO\\ \cite{kaveh2009particle} \end{tabular} & \begin{tabular}{@{}c} MBO\\ \cite{sadollah2012mine} \end{tabular} & \begin{tabular}{@{}c} CBO\\ \cite{kaveh2015comparative} \end{tabular} & \begin{tabular}{@{}c} WCO\\ \cite{sadollah2015water} \end{tabular} & \begin{tabular}{@{}c} DE\\ \cite{ho2016adaptive} \end{tabular} & \begin{tabular}{@{}c} NS\\ \cite{shahabsafa2018novel} \end{tabular} & GA & GSMO\\[0.2em]
\hline
1-4 & 0.196 & 1.800 & 0.196 & 0.196 & 0.196 & 0.196 & 0.196 & 0.141\\[0.2em]
5-12 & 0.563 & 0.602 & 0.563 & 0.563 & 0.563 & 0.563  & 0.563 & 0.563\\[0.2em]
13-16 & 0.442 & 0.111 & 0.391 & 0.391 & 0.391 & 0.391 & 0.391 & 0.391\\[0.2em]
17-18 & 0.563 & 0.111 & 0.563 & 0.563 & 0.563 & 0.563 & 0.563 & 0.563\\[0.2em]
19-22 & 0.563 & 1.266 & 0.563 & 0.563 & 0.563 & 0.563  & 0.563 & 0.563\\[0.2em]
23-30 & 0.563 & 0.563 & 0.563 & 0.563 & 0.563 & 0.563 & 0.563 & 0.563\\[0.2em]
31-34 & 0.111 & 0.111 & 0.111 & 0.111 & 0.111 & 0.111 & 0.111& 0.111\\[0.2em]
35-36 & 0.250 & 0.111 & 0.111 & 0.111 & 0.111 & 0.111 & 0.111 & 0.111\\[0.2em]
37-40 & 1.228 & 0.442 & 1.228 & 1.228 & 1.228 & 1.228 & 1.228 & 1.228\\[0.2em]
41-48 & 0.563 & 0.442 & 0.442 & 0.442 & 0.563 & 0.563 & 0.442 & 0.442\\[0.2em]
49-52 & 0.111 & 0.111 & 0.111 & 0.111 & 0.111 & 0.111 & 0.111 & 0.111\\[0.2em]
53-54 & 0.111 & 0.111 & 0.111 & 0.111 & 0.111 & 0.111 & 0.111 & 0.111\\[0.2em]
55-58 & 1.800 & 0.196 & 1.990 & 1.990 & 1.990 & 1.990 & 1.990 & 1.990\\[0.2em]
59-66 & 0.442 & 0.563 & 0.563 & 0.563 & 0.442 & 0.442 & 0.563 & 0.563\\[0.2em]
67-70 & 0.141 & 0.442 & 0.111 & 0.111 & 0.111 & 0.111 & 0.111 & 0.111\\[0.2em]
71-72 & 0.111 & 0.602 & 0.111 & 0.111 & 0.111 & 0.111 & 0.111 & 0.111\\[0.2em]
\bottomrule
\rule{0pt}{3ex}
\makecell{Optimum \\ mass (lbm)} & 393.38 & 390.73 & 389.33 & 389.33 & 389.33 & 389.33 & 389.33 & \textbf{388.01}\\[0.2em]
\bottomrule[0.1em]
\end{tabular}
\end{center}
\end{table}

\subsection{812-bar lattice structure}
\label{subsec:poisson}
\begin{figure}[!ht]
\centering
    \includegraphics[width=0.8\textwidth]{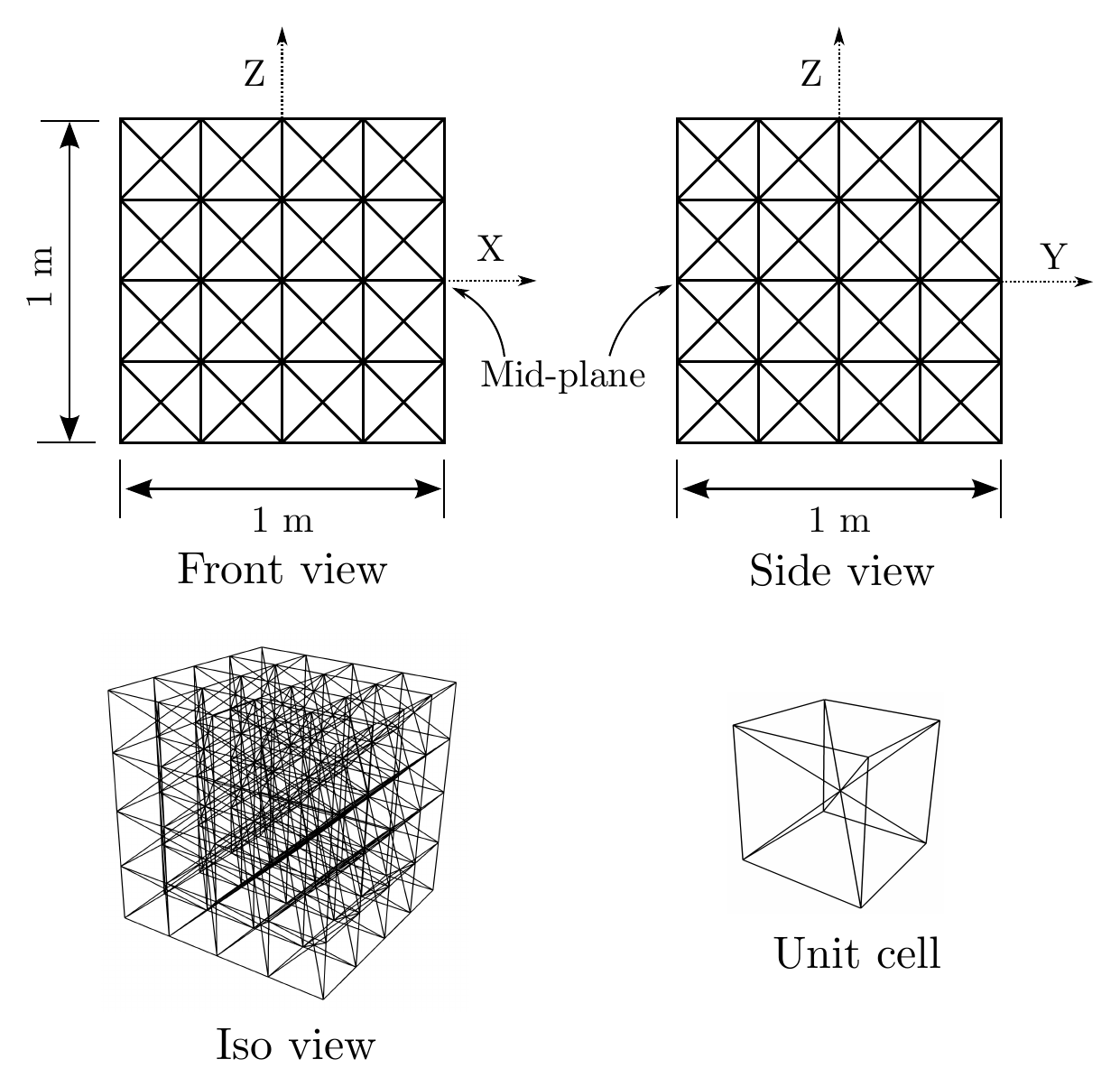}
    \caption{Geometry of the lattice structure}\label{fig:lattice}
\end{figure}

In this example, the goal is to optimize the lattice structure shown in Figure \ref{fig:lattice} to have \emph{nonpositive} Poisson's ratios in X and Y directions. The lattice is symmetric with respect to XY, XZ and YZ planes and is composed of 4 by 4 unit cells depicted in Figure \ref{fig:lattice}. It is pulled along the Z axis by 0.2 m, while the nodes on the outer edges of the mid-plane are expected to satisfy
\begin{equation}\label{eq:midconst}
    \frac{x_{f,i}}{x_{0,i}} \geq 1, \quad \frac{y_{f,i}}{y_{0,i}} \geq 1, \quad i = 1, \cdots, 16,
\end{equation}
where $x_{f,i}$ and $x_{0,i}$ are respectively final and initial $x$-coordinates, and $y_{f,i}$ and $y_{0,i}$ are respectively final and initial $y$-coordinates of Node $i$ on the outer edges of the plane. Equation \ref{eq:midconst} means that the lattice structure must not shrink and possibly expand from the middle in both X and Y directions while being pulled along the Z direction. Note that even though the deformations are large, we still model the structure assuming linear elasticity as the focus of this paper is optimization rather than simulation. In this example, the sole purpose is to satisfy the constraints expressed in \eqref{eq:midconst}, hence the constant function $J = 0$ is considered for the objective function. The lattice is discretized by Euler-Bernoulli beam elements all of which are made of a steel alloy with a Young's modulus of $210$ GPa and a Poisson's ratio of $0.3$.

\begin{figure}[!ht]
\centering
    \includegraphics[width=0.8\textwidth]{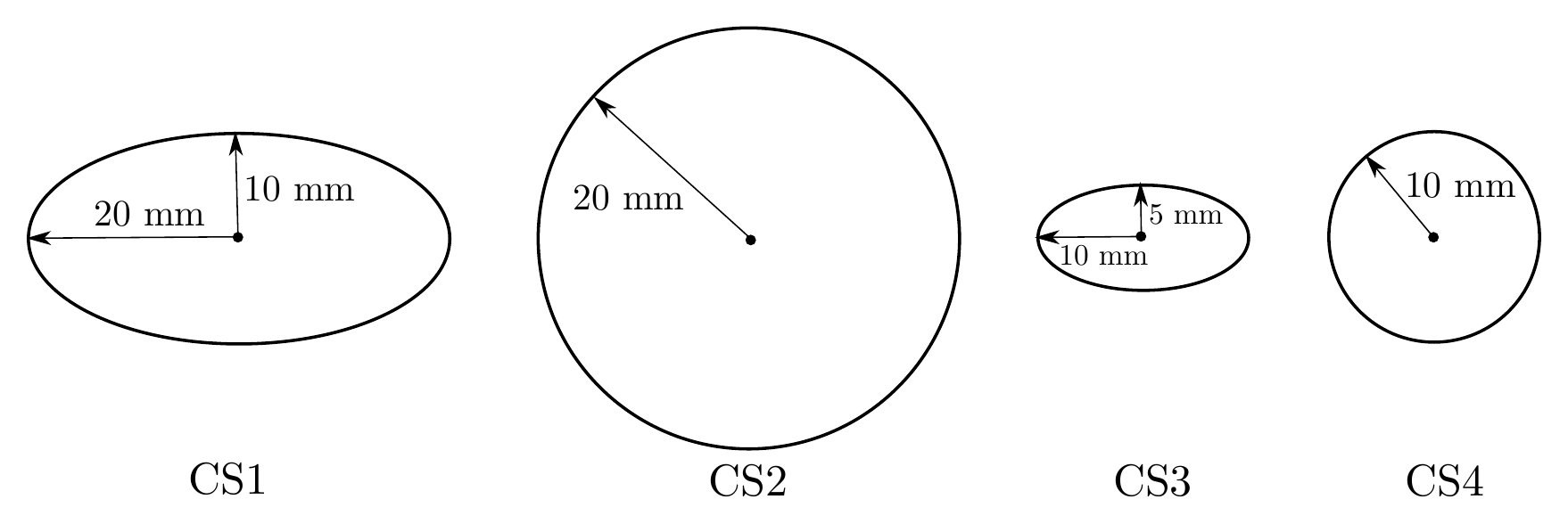}
    \caption{Available cross-sectional choices for the beams in the lattice structure}\label{fig:poissoncs}
\end{figure}

This problem contains a mix of continuous and categorical design variables. The former variables include the orientation of all the 812 beams and the spatial position of the nodes interior to the structure (i.e., those not lying on the lattice's top, bottom, front, back and side planes; 91 nodes in total). This leads to a total of 1085 continuous design variables. For the categorical variables, the 812 beams of the lattice are classified into 4 groups; those along the X, Y and Z axes as well as those along the unit cells' diagonal directions. For each group, 4 cross-sectional choices exist. The details of these cross-sections are provided in Figure \ref{fig:poissoncs}. 

\begin{figure}[ht]
\centering
    \includegraphics[width=\textwidth]{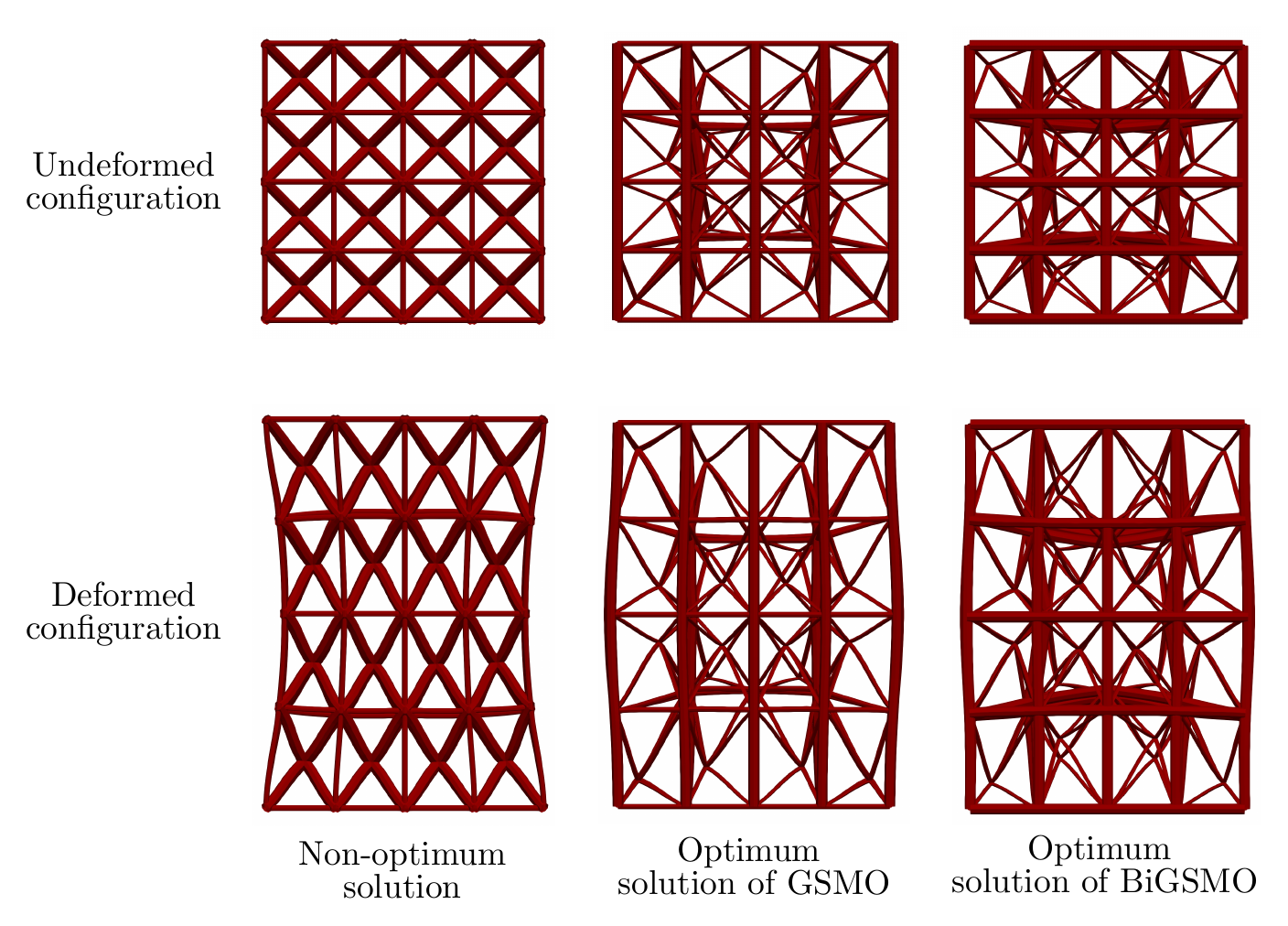}
    \caption{Optimum solutions of the lattice structure due to GSMO and BiGSMO}\label{fig:poissonopt}
\end{figure}

All three methods, GSMO, BiGSMO and GA, were able to produce a feasible solution albeit different from one another. Out of the 10 runs of GSMO and BiGSMO, all converged to the same solution. This once again demonstrates that although a sampling process is involved in these algorithms the optimum solutions generated in different runs are similar (in this example identical). Unlike GSMO and BiGSMO, GA had a hard time finding a feasible solution. For GA, only 3 of the runs led to a solution satisfying the constraints in \eqref{eq:midconst} (All GA runs could find a feasible solution after about 300 iterations). Figure \ref{fig:poissonopt} illustrates the optimum solutions obtained by GSMO and BiGSMO in their undeformed and deformed configurations. The optimum value of cross-sectional choices is provided in Table \ref{table:poscs}.

\begin{table}[ht]
\caption{\label{table:poscs}Optimum cross-sectional choices of the lattice structure}
\begin{center}
\begin{tabular}{cccc}
\toprule[0.1em]
Member group & GSMO & BiGSMO & GA\\[0.2em]
\hline
Along X & CS4 & CS2 & CS1 \\[0.2em]
Along Y & CS4 & CS2 & CS4 \\[0.2em]
Along Z & CS2 & CS1 & CS4 \\[0.2em]
Along diagonal & CS3 & CS3 & CS3 \\[0.2em]
\bottomrule
\end{tabular}
\end{center}
\end{table}

Similar to the previous case study, a major superiority of GSMO and BiGSMO relative to GA is their computational cost. Both GSMO and BiGSMO require only one FE solve per optimization iteration regardless of the number of design variables. For GA on the other hand, since there are 1089 design variables (1085 continuous and 4 categorical variables), with a population size of 10 times the number of design variables, 10890 FE solves are needed per iteration. Even if the FE routine is parallelized with GA, with the computational resource used for the experiment (16 threads), the number of FE solves required for every optimization iteration per CPU core would be $10890/16 \approx 680$. For this example, each FE solve takes about 0.8 seconds on the computer specified earlier, leading to 95 seconds for each GSMO and BiGSMO run (encompassing gradient calculations and other associated overheads). Each GA run with parallelization, on the other hand, extends to around 55000 seconds. Subsequently, the computation cost of GA for this problem is significantly larger than that of GSMO and BiGSMO making it prohibitive for larger-scale problems. The same adversity is expected for other population-based optimization approaches as well. 

\subsection{258-bar bridge structure}
\label{subsec:train}
\begin{figure}[!ht]
\centering
    \includegraphics[width=0.8\textwidth]{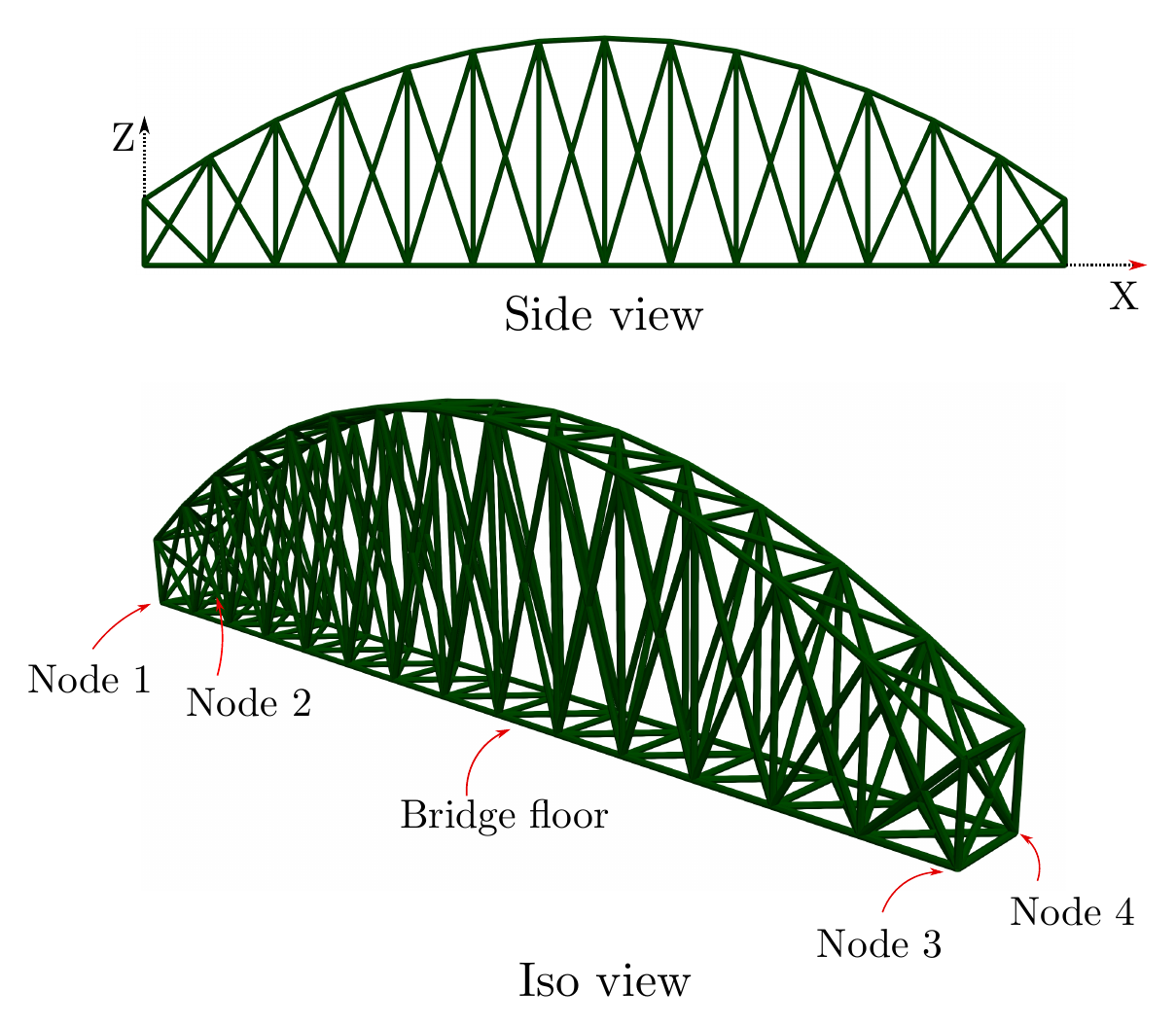}
    \caption{The 258-bar bridge structure}\label{fig:train}
\end{figure}

The bridge structure depicted in Figure \ref{fig:train} is composed of 258 components modeled by Euler-Bernoulli beam elements. The bridge spans 14 m along the X axis, has a width of 1 m and a maximum height of 3.45 m. All of its components are made of steel with a Young's modulus of $210$ GPa, a yield stress of $360$ MPa and a Poisson's ratio of $0.3$. The members lying on the bridge floor are subject to a downward (along the negative Z axis) uniform load of $1000$ N/m. Furthermore, the entire bridge is under the gravity load. The bridge is clamped from Nodes 1 to 4.

\begin{figure}[ht]
\centering
    \includegraphics[width=0.7\textwidth]{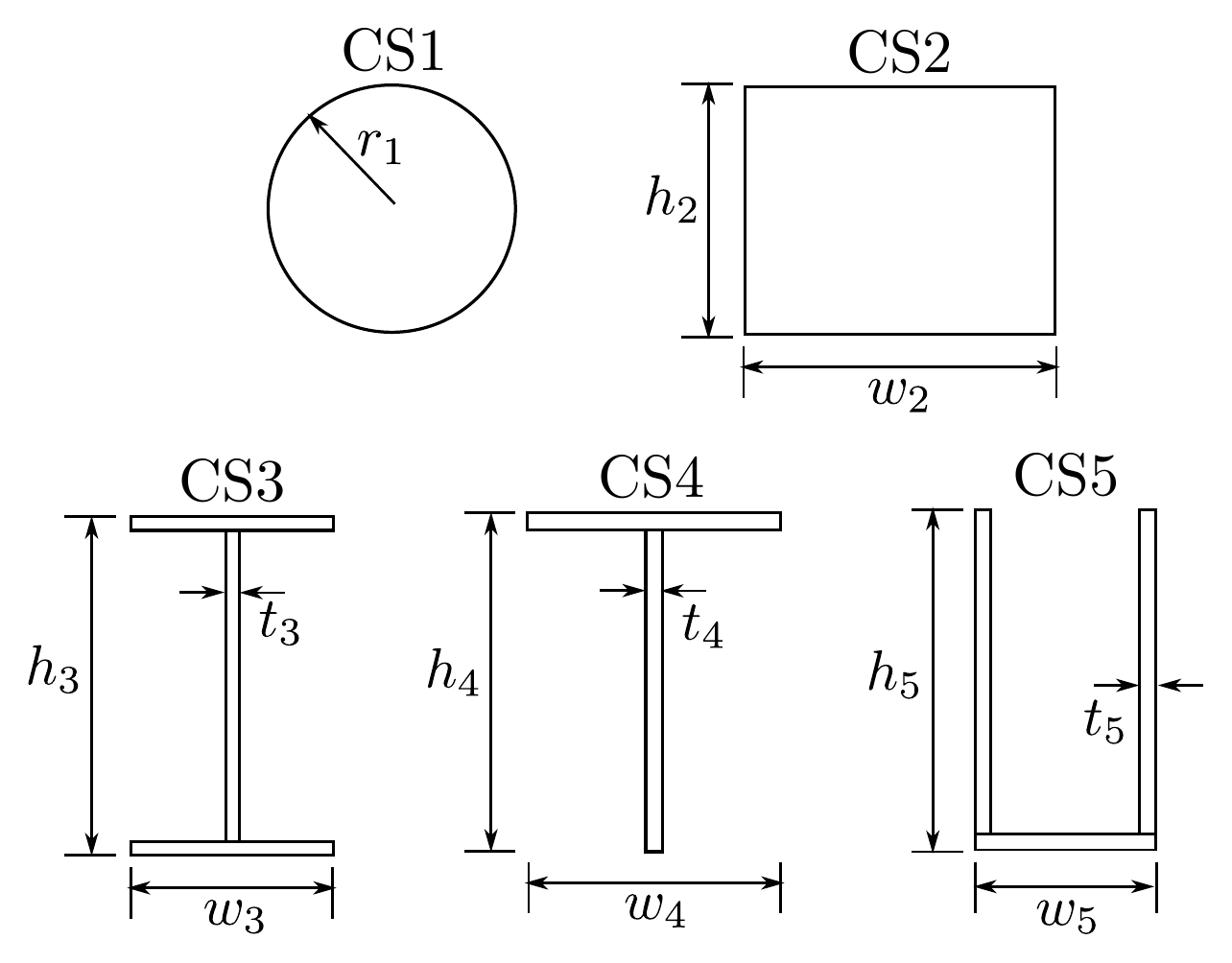}
    \caption{Available cross-sectional choices for the beams in the bridge structure}\label{fig:traincs}
\end{figure}

The objective function of this problem is to minimize the total strain energy of the structure due to the applied loads while the maximum stress in all elements remains below their yield strength. Also, the smallest natural frequency of the structure must be larger than $50$ Hz. Similar to the previous case study, in this problem too we deal with a combination of continuous and categorical design variables. The former variables include the orientation of all 258 beam elements and the length of the beams not lying on the bridge floor (187 beams in total). Hence, there are $445$ continuous design variables in aggregate. The categorical design variables contain the cross-sectional choices of all the elements which are allowed to change independently (i.e., 258 categorical design variables). The available 5 cross-sectional profiles are illustrated in Figure \ref{fig:traincs} and their associated parameters are provided in Table \ref{table:traincs}.

\begin{table}[b]
\caption{\label{table:traincs}Parameter values associated with available cross-sectional choices for the beams in the bridge structure}
\begin{center}
\begin{tabular}{lcccccccccccc}
\toprule[0.1em]
\textbf{Parameter name} & $r_1$ & $h_2$ & $w_2$ & $h_3$ & $w_3$ & $t_3$ & $h_4$ & $w_4$ & $t_4$ & $h_5$ & $w_5$ & $t_5$\\[0.2em]
\textbf{Value (mm)} & 40 & 80 & 100 & 125 & 75 & 5 & 200 & 150 & 10 & 150 & 80 & 7\\[0.2em]
\bottomrule
\end{tabular}
\end{center}
\end{table}

\begin{figure}[ht]
\centering
    \includegraphics[width=0.9\textwidth]{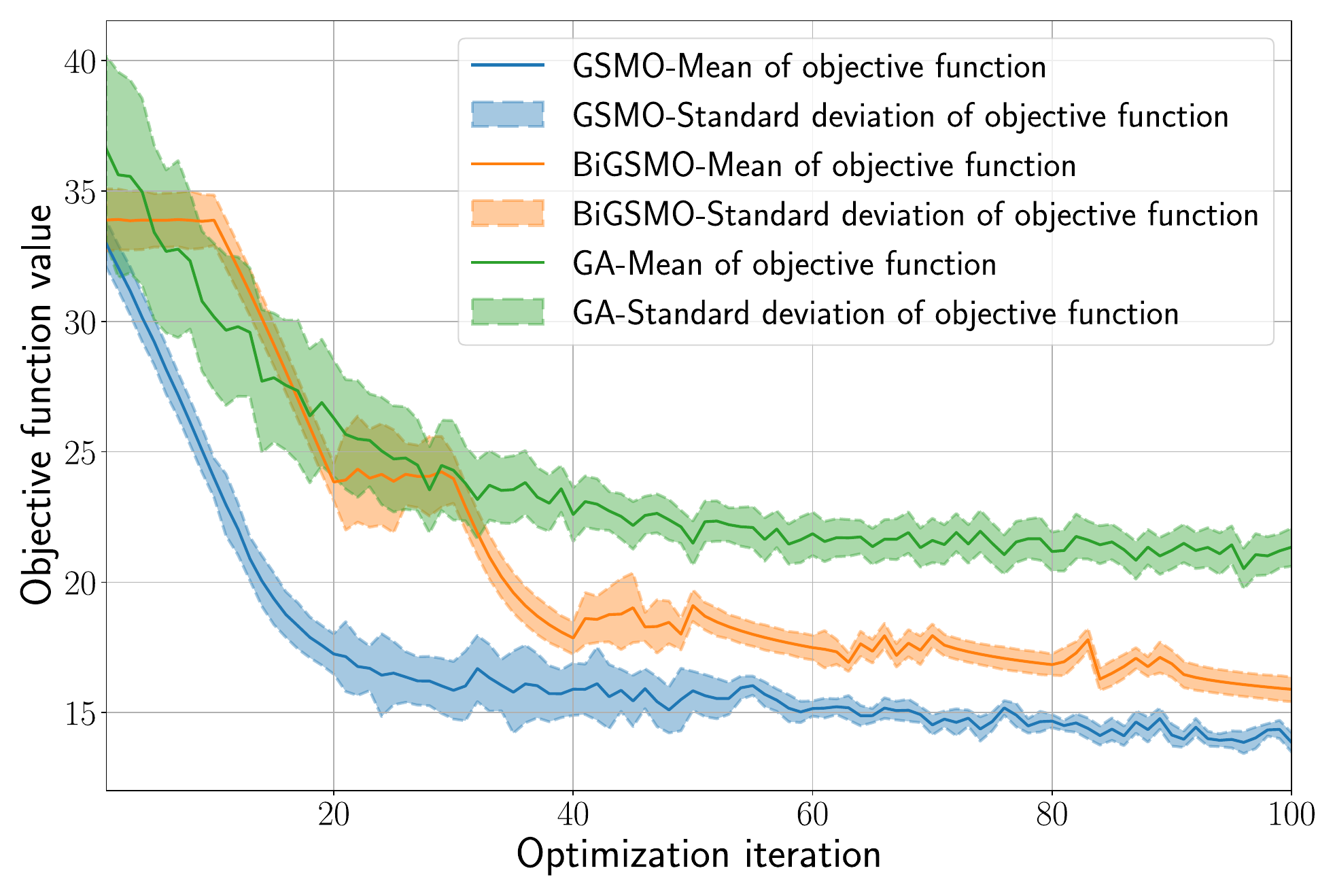}
    \caption{Convergence plot of GSMO, BiGSMO and GA averaged over the 10 runs for the bridge structure problem}\label{fig:trainconv}
\end{figure}

Figure \ref{fig:trainconv} portrays the convergence plot of GSMO, BiGSMO and GA for the 10 runs. Accordingly, GSMO and BiGSMO have a slightly better convergence rate compared to that of GA for this particular problem. both GSMO and BiGSMO exhibit rapid convergence during the initial optimization stages, followed by a gradual deceleration as they approach a local optimum. This slowdown in convergence can be attributed to the diminishing gradient magnitude as the optimization nears a local optimum, leading to smaller solution updates at each iteration. Furthermore, as perhaps expected, two distinct stages can be observed in the BiGSMO plot; a somewhat flat region followed by a steep decline in every 10 iterations. This behavior is due to the fact that in BiGSMO the categorical and continuous design variables are handled in different levels of the optimization routine. As detailed in Algorithm \ref{alg:bioptscheme}, considering each 20-iteration interval, the first 10 iterations are dedicated to evolving the categorical variables and the second 10 iterations advance the continuous variables.

The best objective value among the 10 runs as well as their average and standard deviations for the three optimization methods are presented in Table \ref{table:trainres}. Accordingly, GSMO outperforms BiGSMO and GA in all three aspects. The advantage of GSMO over GA for this example may be mainly attributed to the existence of a considerably large number of design variables, particularly the categorical variables. Since GSMO relies on sensitivity analysis, it is able to explore the high-dimensional design space more effectively compared to GA. Therefore, although GA is by construction capable of finding the global optimum, it falls short of finding one in this problem. On the other hand, the best solution found via BiGSMO is not as good as those of GSMO and GA. The reason could be due to the bilevel nature of BiGSMO. Since the categorical and continuous variables are handled in two different stages in this algorithm, at each stage, a portion of the design space is obscured and inadmissible for exploration by the optimizer. For this example, this has led to a less qualified solution compared to that generated by the other two methods. 

Other noteworthy considerations in Table \ref{table:trainres} are due to average and standard deviation values. Unlike the previous case study, for this problem, the optimum solutions produced in the 10 runs using GSMO and BiGSMO are not unique, pointing to the fact that the sampling involved in these two approaches may lead to generating different solutions every time they are run. However, as can be seen, the average values are quite close to their corresponding best optimum values and the solutions are highly clustered around the best solutions. In this problem, that is not the case for GA. Looking at GA's average and standard deviation values, there is a high chance of getting a solution relatively far from the best possible solution every time GA is executed. While this trait can sometimes help in obtaining hard-to-find global optimums, such is not the case in this experiment as GA failed to find the best solution compared to GSMO. This is in fact a well-known phenomenon for other population-based optimizers as well. 

\begin{table}[ht]
\caption{\label{table:trainres}Performance of GSMO, BiGSMO and GA on the bridge structure}
\begin{center}
\begin{tabular}{ccccc}
\toprule[0.1em]
Method & \begin{tabular}{@{}c} Best Value\\(N.m) \end{tabular}   & \begin{tabular}{@{}c} Average\\(N.m) \end{tabular}  & \begin{tabular}{@{}c} Standard Deviation\\(N.m) \end{tabular} & \begin{tabular}{@{}c} Approximate Execution Time\\ (s) \end{tabular}\\[0.2em]
\hline
GSMO & 12.998 & 13.643 & 0.348 & 140 \\[0.2em]
BiGSMO & 15.121 & 15.853 & 0.350 & 140 \\[0.2em]
GA & 13.642 & 20.520 & 2.712 & 49000 \\[0.2em]
\bottomrule
\end{tabular}
\end{center}
\end{table}

\begin{figure}[ht]
\centering
    \includegraphics[width=1\textwidth]{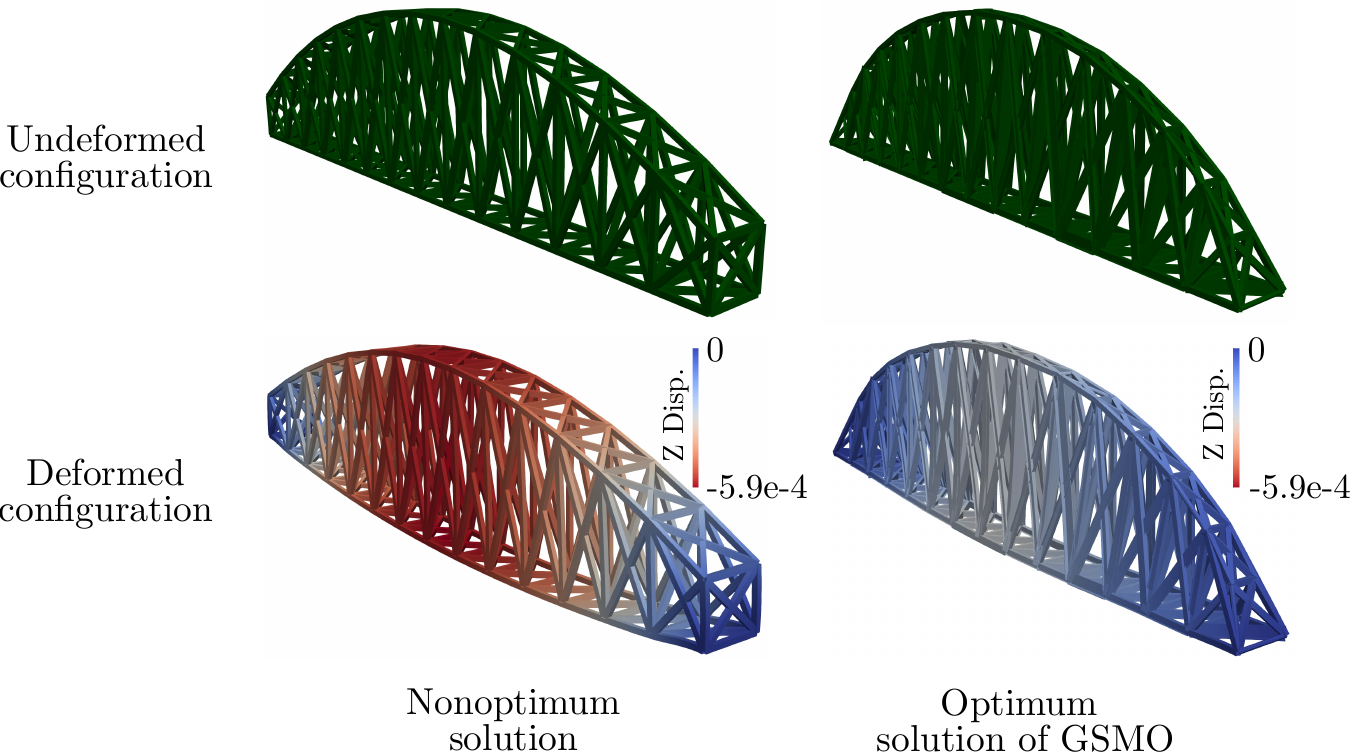}
    \caption{Optimum solution of the bridge structure due to GSMO. The deformations are scaled by a factor of 1000.}\label{fig:traindef}
\end{figure}

Similar to other case studies, the advantage of GSMO and BiGSMO over GA can be appreciated more from the computational perspective. For this example, the computational cost per iteration of GA is about 7000 times larger than that of both GSMO and BiGSMO. The difference once again is due to requiring one linear elasticity and one modal analysis solve per optimization iteration for GSMO and BiGSMO as compared to 7030 of each solve for GA (total number of design variables is $445 + 258 = 703$). For this problem, one FE linear elasticity solve and one FE modal analysis solve combined take about 1.1 seconds on the computer specified earlier. This leads to approximately 140 seconds per run on average (including other overloads such as gradient calculations) for both GSMO and BiGSMO and about 49000 seconds per run (executing 16 solves in parallel per iteration) for GA. Figure \ref{fig:traindef} shows the GSMO's best optimum solution and nonoptimum (initial) solution in their undeformed and deformed configurations.

\section{Conclusion}
\label{sec:conclusion}

The optimal design of real-world frame structures presents a significant challenge due to the presence of a large number of categorical design variables, such as cross-sectional profiles and material choices. The categorical nature of these variables precludes the employment of gradient-based optimizers and necessitates the use of gradient-free optimizers that are known to be less efficient. To address this challenge, this paper proposes a gradient-based optimizer that leverages the GSM method. The GSM method represents categorical choices as continuous probability distributions and enables the sampling process from such distributions to be differentiable. This in turn allows computing the sensitivities of objective and constraint functions with respect to the categorical design variables. This information can be combined with the sensitivities with respect to the continuous design variables to enable either simultaneous or bilevel optimization of categorical and continuous design variables, corresponding to the development of GSMO and BiGSMO, respectively.

Relative to GA, we have demonstrated that both GSMO and BiGSMO can find optimal solutions in a significantly shorter amount of time because the number of FE solves required is orders of magnitudes smaller. Specifically, our case studies showed that compared to GA, the computational costs of GSMO and BiGSMO were $\mathcal{O}(10^3)$ lower for the lattice and bridge structure problems. This remarkable improvement can be ascribed to two primary factors. Firstly, we have essentially transformed combinatorial optimization problems involving categorical design variables---which typically have exponential complexity---into problems involving only continuous design variables with a polynomial complexity \cite{LENSTRA1979121}. Secondly, the incorporation of continuous design variables enables the application of gradient-based optimizers, which are well-known for their superior efficiency and scalability in comparison to gradient-free optimization methods \cite{mdobook}.

In addition to the computational advantages, our case studies also showed that GSMO and BiGSMO found better solutions with more consistency than GA. In the 72-bar truss structure, GSMO generated an optimal solution that outperformed those of other methodologies reported in various studies. For the 812-bar lattice structure, both GSMO and BiGSMO consistently found a unique optimum solution in all 10 runs, while GA only managed to find an optimum solution in 3 out of the 10 runs. Lastly, in the 258-bar bridge structure, both GSMO and BiGSMO resulted in a small standard deviation in optimal objective function values across the 10 runs, with GSMO delivering a noticeably better solution compared to BiGSMO and GA.

These findings were somewhat surprising because GA is known to be more adept at finding global optimum solutions than gradient-based approaches due to its ability to explore the solution space, albeit at the expense of efficiency. We hypothesize several reasons for this outcome. Assuming the general understanding that the gradient-based optimizers are susceptible to falling into local optima, it is likely that the consistent sets of solutions found by our method were indeed local minima. Conversely, because of the high complexity of frame structure design problems, GA was unable to get close to any optimal solution (neither local nor global) within the given number of iterations. Hence, the best solutions found by GA in our case studies tended to have high variance and subpar objective values. Had we run GA for a longer period of time (as was done in the second case study), it might have been possible that GA would have eventually converged to a better optimal solution. However, we conjecture that this would require an extraordinary amount of computational time for solving real-world design problems with even greater complexity than those studied in this paper.

Another important advantage of our method is that we are able to simultaneously optimize both categorical and continuous design variables. Typically, the problems involving both variable types are solved in a bilevel manner, wherein a gradient-free optimizer is used to solve for categorical variables at the outer level while a gradient-based optimizer is used to solve for continuous variables at the inner level. In this paper, we have investigated how this bilevel approach performs using a gradient-based optimizer for both levels. The last case study of bridge structure design demonstrated that the simultaneous approach employed by GSMO outperforms the bilevel approach employed by BiGSMO. This superior performance can likely be attributed to GSMO's enhanced efficacy in exploring the design space by considering both categorical and continuous design variables concurrently, while BiGSMO solves for each variable type sequentially.

Regarding the limitations, a distinctive aspect of our method is the requirement for each categorical variable to be represented by a set of continuous attributes (properties) to compute sensitivities. For instance, computing the gradients of an objective function with respect to cross-sectional profiles requires the gradients of the objective function with respect to the cross-sectional areas and moments of inertia, as expressed in \eqref{eq:djdsfinal}. In certain applications, such characterization of categorical variables might not be readily available (e.g., the choice of joint types for a multi-component structural problem). Nonetheless, we assert that the majority of categorical choices involved in engineering design problems are associated with continuous parameters, enabling sensitivity computation and making our approach applicable to a wide range of applications.

Another limitation is that since GSMO and BiGSMO are gradient-based approaches, they are susceptible to getting trapped in local minima. However, the inherent stochastic nature of the GSM method provides a means to control the exploration capability of our method, thereby increasing the likelihood of finding global optima. Specifically, the Gumbel temperature annealing schedule can be varied such that the optimizer can use more iterations to sample a variety of solutions before converging. In addition, one can take the initial and final Gumbel temperatures as design variables and determine their optimum values for a given problem alongside other design variables. Furthermore, adopting a multi-start strategy where the problem is run using various initial solutions might mitigate the risk of being trapped in a local minimum. Such extensions should be explored in future work.

Furthermore, our broader application experience beyond the cases showcased in this study indicates that as the number of choices for individual categorical variables increases, both GSMO and BiGSMO exhibit oscillatory convergence behavior and, occasionally, may even fail to converge to a mathematically optimal solution. This phenomenon stems from the fact that when numerous choices are available for a variable, generating a sole sample from the distribution might not adequately capture the actual distribution associated with that variable. A plausible resolution could involve generating multiple samples for each categorical variable and subsequently selecting the sample generated most frequently. Lower Gumbel temperatures can also reduce the variance of the generated samples. 

Lastly, in the presented second and third case studies, we compared our method to only one specific derivative-free method, GA. While GA is a well-known method that has been successfully applied to various engineering design applications, there are other derivative-free methods suitable for solving problems with categorical variables such as Tabu search, Monte Carlo tree search and estimation of distribution algorithms. However, all of these methods require a large number of function evaluations as they rely on either unguided (i.e., not guided by gradients) search moves or population-based improvements. Therefore, we are confident that the proposed method in this study will outperform such alternatives in terms of computational efficiency. 

It is important to note that while we developed GSMO and BiGSMO in this study with structures undergoing linear elasticity behavior in mind, the algorithms presented in Algorithms \ref{alg:optscheme} and \ref{alg:bioptscheme} can be adapted to any optimization problem with mixed categorical and continuous design variables subject to other types of governing equations, with only minor modifications. The core principles of the presented algorithms remain unchanged; only the adjoint equations need to be adjusted accordingly. This versatility further underscores the potential of GSMO (and BiGSMO) for a wide range of applications, including structures with nonlinear behaviors.

\newpage
\begin{appendices}

\section{}
\label{appen:gm}
\begin{prop}\label{pr:gm}
The samples generated through \eqref{eq:gsm} have the same probability distribution as $\boldsymbol{\theta}$.
\end{prop}

\begin{proof}
Let us first introduce the standard Gumbel distribution Gumbel(0, 1). Its cumulative distribution function $F$ and probability density function $f$ read
\begin{equation}
    F_X(x) = P (X \leq x) = \exp\left( -\exp(-x) \right), \quad f_X(x) = \frac{d}{dx} F_X(x) =\exp\left(-(x+\exp(-x)) \right),
\end{equation}
where the subscript $X$ is the random variable associated with Gumbel(0, 1). Defining $Y$ and $\Theta$ as two other random variables satisfying $Y = \Theta + X$, we have
\begin{equation}\label{eq:cdf}
    F_Y(\overline{y}) = P (Y \leq \overline{y}) = P ( \Theta + X \leq \overline{y}) = P (X \leq \overline{y} - \Theta) = F_X(\overline{y} - \Theta) = \exp\left( -\exp(\Theta - \overline{y}) \right),
\end{equation}
leading to
\begin{equation}\label{eq:pdf}
    f_Y(\overline{y}) = \frac{d}{d\overline{y}}F_Y(\overline{y}) = \frac{d}{d\overline{y}} \exp\left( -\exp(\Theta - \overline{y}) \right) = \exp \left( \Theta - \overline{y} \right) \exp\left( -\exp(\Theta - \overline{y}) \right).
\end{equation}
Now, defining $z_i := \theta_i + G^{(i)}$ as in \eqref{eq:gsm}, to prove the proposition, it suffices to show that the probability of $z_k, \: k \in [1,N]$ being the maximum of set $\{z_1, \cdots, z_N \}$ is $[ \text{softmax} (\boldsymbol{\theta}) ]_k$ as given in \eqref{eq:thetatop}. In other words,
\begin{equation}\label{eq:prop1}
    P(z_k = \max \{z_1, \cdots, z_N \}) = \frac{\exp(\theta_{k})}{\sum_{i=1}^{N} \exp(\theta_{i})}, \quad k \in [1,N].
\end{equation}
We proceed as follows: Another way of interpreting $z_k = \max \{z_1, \cdots, z_N \}$ is to set $\overline{z} := z_k$ and write $z_k = \max \{z_1, \cdots, z_N \}$ as
\begin{equation}
    z_1 \leq \overline{z}, \cdots, z_{k-1} \leq \overline{z}, z_k = \overline{z}, z_{k+1} \leq \overline{z}, \cdots, z_N \leq \overline{z}.
\end{equation}
Then, since the samples $z_i$ are generated independently, using \eqref{eq:cdf} and \eqref{eq:pdf} we have
\begin{equation}
    \begin{aligned}
    P \left(z_k = \max \{z_1, \cdots, z_N \} \right) &= P \left( z_1 \leq \overline{z}, \cdots, z_{k-1} \leq \overline{z}, z_k = \overline{z}, z_{k+1} \leq \overline{z}, \cdots, z_N \leq \overline{z} \right) \\
    &= \int_{-\infty}^{\infty} \left[ \exp \left( \theta_k - \overline{z} \right) \exp\left( -\exp(\theta_k - \overline{z}) \right)  \prod_{i=1, i \neq k }^N \exp\left( -\exp(\theta_i - \overline{z}) \right) \right] d\overline{z}\\
    &= \int_{-\infty}^{\infty} \left[ \exp \left( \theta_k - \overline{z} \right) \prod_{i=1 }^N \exp\left( -\exp(\theta_i - \overline{z}) \right) \right] d\overline{z}\\
    &= \int_{-\infty}^{\infty} \left[ \exp \left( \theta_k - \overline{z} \right) \exp \left(-\exp \left( -\overline{z}\right) \sum_{i=1}^N \exp(\theta_i)  \right) \right] d\overline{z}\\
    &= \frac{\exp(\theta_k) \exp \left(-\exp \left( -\overline{z}\right) \sum_{i=1}^N \exp(\theta_i)  \right)}{\sum_{i=1}^N \exp(\theta_i)} \Bigg|_{-\infty}^{\infty} = \frac{\exp(\theta_k)}{\sum_{i=1}^N \exp(\theta_i)}.
    \end{aligned}
\end{equation}
Therefore, \eqref{eq:prop1} holds and the proposition is proved.
\end{proof}

\newpage
\section{}
\label{appen:bigsmo}
\vspace{-0em}
\begin{algorithm}[!htb]
\caption{The BiGSMO scheme}\label{alg:bioptscheme}
\SetAlgoLined
\KwInput{Objective function $J$, constraint functions $\boldsymbol{g}$, continuous design variables $\boldsymbol{x}$ and their bounds, categorical design variables $\boldsymbol{c}$ and their available choices, GSM annealing scheme.}
\KwOutput{Optimum design.}
    Initialize $\boldsymbol{x}$\;
    Initialize $\boldsymbol{\theta}_i, \: i=1,\cdots,n_c,$ for each categorical variable $c_i$ by entry values of zero\;
    \While{outer optimizer not converged}
    {\label{ln:outer}
        \While{inner optimizer not converged}
        {\label{ln:inner}
            \For{$i=1, \cdots, n_c$}{
                Generate Gumbel noises $G^{(j)}, \:j=1,\cdots,N_i,$ from $\text{Gumbel}(0,1)$\;
                Compute $(\widetilde{\boldsymbol{s}}_{i})_j, \:j=1,\cdots,N_i,$ using \eqref{eq:zvec} and $\nabla_{\boldsymbol{\theta}_i} \widetilde{\boldsymbol{s}}_i$ via \eqref{eq:dz}\;\label{st:bidsdtheta}
                Calculate $\widehat{\boldsymbol{s}}_i$ through \eqref{eq:yvec}\;
            }
            Solve the governing equations $\mathbf{K} \boldsymbol{u} = \boldsymbol{f}$ using $\widehat{\boldsymbol{s}}_i, \: i=1, \cdots, n_c,$\ and get $\boldsymbol{u}$\;
            Compute $J$ and $\boldsymbol{g}$ values using $\widehat{\boldsymbol{s}}_i, \: i=1, \cdots, n_c,$\ and $\boldsymbol{u}$\;
            Calculate $\nabla_{\boldsymbol{u}} J$ then find $\boldsymbol{\lambda}_J$ by solving \eqref{eq:adjoint}\;\label{st:biadjj}
            Compute $\nabla_{\boldsymbol{u}} g_i, \: i=1,\cdots, n_g,$ then find $\boldsymbol{\lambda}_{g_i}, \: i=1,\cdots, n_g,$ by solving \eqref{eq:adjointg}\;\label{st:biadjg}
            \For{$i=1, \cdots, n_c$}{
                Form the attribute matrix $\mathbf{A}_i$ as in \eqref{eq:attmat}\;
                Calculate $\partial J / \partial \boldsymbol{a}_i$, $\partial \boldsymbol{g} / \partial \boldsymbol{a}_i$, $\partial \mathbf{K} / \partial \boldsymbol{a}_i$ and $\partial \boldsymbol{f} / \partial \boldsymbol{a}_i$ using the attributes of the selected class for this categorical variable\;
                Compute $\nabla_{\boldsymbol{a}_i} J$ and $\nabla_{\boldsymbol{a}_i} \boldsymbol{g}$ utilizing the adjoint vectors found in Steps \ref{st:biadjj} and \ref{st:biadjg}, $\nabla_{\boldsymbol{\theta}_i} \widetilde{\boldsymbol{s}}_i$ found in Step \ref{st:bidsdtheta} and employing \eqref{eq:djdai}\;
                Get $\nabla_{\boldsymbol{\theta}_i} J$ and $\nabla_{\boldsymbol{\theta}_i} \boldsymbol{g}$ through \eqref{eq:djdsfinal}\;\label{st:bisentheta}
            }
            Update $\boldsymbol{\theta}_i, \: i=1,\cdots,n_c$, using the sensitivities found in Step \ref{st:bisentheta}\;
        }
        Update $\widetilde{\boldsymbol{s}}_i$ and $\widehat{\boldsymbol{s}}_i, \: i=1,\cdots,n_c,$ using the new $\boldsymbol{\theta}_i$ values\; \label{st:bidnews}
        Solve the governing equations $\mathbf{K} \boldsymbol{u} = \boldsymbol{f}$ using $\widehat{\boldsymbol{s}}_i, \: i=1, \cdots, n_c,$ in Step \ref{st:bidnews} and get $\boldsymbol{u}$\;
        Compute the new values of $J$, $\nabla_{\boldsymbol{u}} J$, $\boldsymbol{\lambda}_J$; and $g_i$, $\nabla_{\boldsymbol{u}} g_i$ and $\boldsymbol{\lambda}_{g_i}, \: i=1,\cdots, n_g$\;
        Calculate $\partial J / \partial x_i$, $\partial \boldsymbol{g} / \partial x_i$, $\partial \mathbf{K} / \partial x_i$ and $\partial \boldsymbol{f} / \partial x_i, \: i=1,\cdots,n_x$\;
        Get $\nabla_{\boldsymbol{x}} J$ and $\nabla_{\boldsymbol{x}} \boldsymbol{g}$ incorporating the corresponding adjoint vectors and \eqref{eq:sensjxfinal}\;\label{st:bisenx}
        Update $\boldsymbol{x}$ using the sensitivities found in Steps \ref{st:bisenx}\;
    }
\end{algorithm}

\end{appendices}

\newpage
\bibliography{main}
\bibliographystyle{ieeetr}
\end{document}